\documentclass[pra,aps,twocolumn,superscriptaddress,floatfix,showpacs]{revtex4-1}

\usepackage{graphicx}
\usepackage{amsmath}
\usepackage{amssymb}
\usepackage{wasysym}

\newcommand{\bra}[1]{\langle #1|}
\newcommand{\ket}[1]{|#1\rangle}
\newcommand{\expectation}[1]{\left\langle #1\right\rangle}
\newcommand{\inner}[2]{\langle #1 | #2 \rangle}
\newcommand{\rdm}{\rho}

\newcommand{\lambdathermal}{\lambda_\text{th}}
\newcommand{\lambdaeth}{\lambda_\text{ETH}}
\newcommand{\Uthermal}{U_{\rm th}}
\newcommand{\Teff}{T_{\text{eff}}}
\newcommand{\Tr}{\text{Tr}}

\newcommand{\lambdanp}{\lambda_\text{np}}
\newcommand{\EP}{{\text{EP}}}
\newcommand{\spreadproj}{\sigma_\EP}
\newcommand{\pldos}{{\text{(2)}}}
\newcommand{\nparticle}{n}
\newcommand{\Nparticle}{N}
\newcommand{\nstate}{M}
\newcommand{\egysys}{\varepsilon}
\newcommand{\egybath}{\epsilon}

\begin{document}

\title{Thermalisation of Local Observables in Small Hubbard Lattices}

\author{S.~Genway}\address{School of Physics and Astronomy, The
  University of Nottingham, Nottingham NG7 2RD, United Kingdom}

\author{A.~F.~Ho}\address{Department of Physics, Royal Holloway
  University of London, Egham, Surrey TW20 0EX, United Kingdom}

\author{D.~K.~K.~Lee}\address{Blackett Laboratory, Imperial
  College London, London SW7 2AZ, United Kingdom}


\date{\today}

\begin{abstract} 
  We present a study of thermalisation of a small isolated Hubbard lattice
  cluster prepared in a pure state with a well-defined energy.  We
  examine how a two-site subsystem of the lattice thermalises with the
  rest of the system as its environment.  We explore numerically the
  existence of thermalisation over a range of system parameters,
  such as the interaction strength, system size and the strength of
  the coupling between the subsystem and the rest of the lattice.  We
  find thermalisation over a wide range of parameters and that 
  interactions are crucial for efficient thermalisation of small systems.  We
  relate this thermalisation behaviour to the eigenstate
  thermalisation hypothesis and quantify numerically the extent to
  which eigenstate thermalisation holds. We also verify
  our numerical results theoretically with
  the help of previously established results from random matrix theory
  for the local density of states, particularly the finite-size
  scaling for the onset of thermalisation.
\end{abstract}

\maketitle

\section{Introduction}

Understanding the quantum origins of statistical mechanics has seen
renewed interest over the last few years, in part motivated by
experimental progress in degenerate atomic gases, but
also due to independent theoretical advances~\cite{Cazalilla2010,Polkovnikov2011,Yukalov2011,QuantumThermodynamics}. The central question is as
follows. Consider a closed quantum system prepared in a pure quantum
state. Does it evolve in time to a thermal state? If so, in what sense
is it a thermal state?  

\begin{figure}[bht]
\includegraphics[scale=0.5]{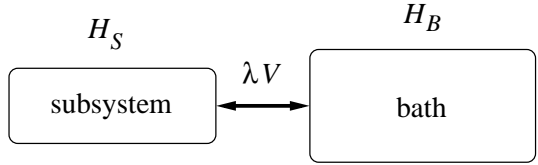}
\caption{Schematic diagram of a closed system divided conceptually 
  into a subsystem and a bath.}
\label{sysbath}
\end{figure}

In this paper, we focus on observables that are local to a subsystem
of the full system. Thus, we discuss the `thermalisation' of this
subsystem with the rest of the system as a bath (see
Fig.~\ref{sysbath}). 
We will discuss the conditions for the eventual thermalisation of this
subsystem. This has been studied in many systems~\cite{Saito1996,Yuan2009,*Jin2010,Henrich2005,Gogolin2011,Cho2010}
and we will study a system of interacting fermions in this context.
The picture of a local subsystem in a closed system also naturally
makes contact with the conventional framework of statistical mechanics
where thermal equilibrium is achieved by a weak coupling $\lambda V$
between a system and its environment.

Thermalisation in closed quantum systems has been shown to have its
origins in entanglement. 
To be specific, let us consider a composite system with a
Hamiltonian $H = H_{S} + H_{B} + \lambda V$ where $H_{S,B}$ describes
the dynamics of subsystem ($S$) and the bath ($B$) respectively while
$\lambda V$ couples the subsystem to the bath. The exact eigenstates
$\ket{A}$ of this Hamiltonian are typically superpositions of many
eigenstates of the decoupled system ($\lambda=0$) which are products
of the subsystem and bath states: $\ket{A} = \sum_{sb}c_{sb}
\ket{s}_S\otimes \ket{b}_B$.
The idea of `canonical
typicality'~\cite{Tasaki1998,Goldstein2006,Popescu2006,
  Reimann2007,*Reimann2008,Linden2009} states that almost any pure
state composed of many energy eigenstates $\ket{A}$ within a narrow
energy window will give rise to a canonical distribution for the
measurements of local or few-body observables within the subsystem.
This emerges because the pure state is an entangled combination of
subsystem and bath eigenstates. We will consider in this paper a system
prepared initially at time $t=0$ in a pure state that is a product
state of the subsystem and bath states.  Such a state is typically a
superposition of many closed-system eigenstates: $\ket{\Psi(t=0)} =
\ket{\phi}_S\otimes \ket{\psi}_B = \sum_A d_A(t=0) \ket{A}$. While
this initial state is special and cannot be considered as `typical',
we expect the wavefunction will, in general, evolve in time
($\ket{\Psi(t)} = e^{-iHt}\ket{\Psi(0)}$) towards a state that falls
into the domain where canonical typicality applies.  The sufficient
conditions for this to occur have discussed in recent
papers~\cite{Linden2009,Gogolin2010,Lychkovskiy2010}. In this paper,
we will investigate conditions for thermalisation in a small
Hubbard-model system.

An alternative view is the eigenstate thermalisation
hypothesis~\cite{Deutsch1991,Srednicki1994,Rigol2008} (ETH).  The time
evolution of any few-body observable $\bra{\Psi(t)}O\ket{\Psi(t)}$
involves the interference of eigenstates at different frequencies:
$\expectation{O} = \sum_{AB} d^*_Ad_B \bra{A}O\ket{B} e^{i(E_A-E_B)t}$
where $E_{A,B}$ is the energy of the eigenstates $\ket{A}$ and
$\ket{B}$. The eigenstate thermalisation hypothesis says that
destructive interference removes all $A\neq B$ terms and that
\begin{equation}
\langle A | O | A \rangle \approx \langle O
\rangle_{E_A}
\label{eq:observable_eth}
\end{equation}
where the right-hand side denotes the thermal average of $O$
when the total system has energy $E_A$. This paints a very different
picture of thermalisation compared to the scenario for classical statistical
mechanics where states diffuse ergodically through phase space
constrained by energy conservation.

These concepts are powerful because they guarantee thermalisation for
closed quantum systems. They depend crucially on the very high
dimensionality of the Hilbert space of quantum states. In this paper,
we aim to gain insight into these ideas by testing the limits of these
hypotheses in terms of the breakdown of thermalisation for a
\emph{small} closed quantum system.  We take our motivation from cold
atom experiments with optical lattices and single-site
addressibility~\cite{Sherson2010,Bakr2010}.  We choose a lattice
system of interacting fermions in a normal metallic state.  In
particular, we present a study of the thermalisation of a composite
system consisting of a 2-site subsystem and a $(L-2)$-site bath in a
one-dimensional Hubbard ring. We avoid the issue of integrability and
the generalised Gibbs ensemble~\cite{Rigol2007} by choosing parameters
such that it is a non-integrable system.

In order to study the thermalisation of the subsystem, 
we need to calculate the long-time
behaviour of reduced density matrix $\rdm$ of the subsystem:
\begin{equation}
\rdm(t)= \text{Tr}_B \ket{\Psi(t)}\bra{\Psi(t)}\,.
\label{eq:rdmdef}
\end{equation}
where Tr$_B$ denotes a trace over the bath degrees of freedom.  A
thermalised system corresponds to a diagonal reduced density matrix
with diagonal elements given by the Gibbs distribution.  We explore
whether thermalisation occurs over a range of system parameters. We
find numerically (section \ref{sec:numerics}) that thermalisation
occurs in surprisingly small systems. For a system of a given size,
there is a threshold for the onset of thermalisation, both in terms of
the coupling strength $\lambda$ and the interaction strength. In
particular, we study the size dependence of the threshold
$\lambdathermal$ that the coupling strength has to exceed to achieve
thermalisation.  We demonstrate that this threshold for thermalisation
agrees with the ETH criterion \eqref{eq:observable_eth} for
thermalisation (sections \ref{sec:eigentherm}).  Indeed, a theoretical
threshold $\lambdaeth$ determined from the ETH criterion has the same
size dependence as the empirical $\lambdathermal$ (section
\ref{sec:SystemSizeScalingInteracting}).  We also argue that both of
these thresholds mark the onset of non-perturbative mixing of
eigenstates due to the subsystem-bath coupling (at a threshold
$\lambdanp$).

From a separate perspective~\cite{Deutsch1991,Srednicki1994}, we can
study the thermalisation process in terms of the statistics of the
eigenstates. We study the statistics of the overlap $\inner{A}{sb}$ of
the eigenstates $\ket{A}$ of the coupled system with the eigenstates
of the decoupled system which are product states $\ket{sb}\equiv
\ket{s}_S\otimes \ket{b}_B$. Interestingly, at weak subsystem-bath
coupling where the onset of thermalisation occurs, the distribution
for the overlaps fits a hyperbolic secant distribution (section
\ref{sec:ldos}). This is in contrast to previous
conjectures~\cite{Berry1977,Deutsch1991,Srednicki1994} from random
matrix theory which suggest that these types of overlaps should follow
a Normal distribution at weak coupling.

Using our results for the overlap distribution, we
show numerically (section \ref{sec:eigentherm}) that the eigenstate
thermalisation hypothesis \eqref{eq:observable_eth} holds for the
projection operator $P_s = \sum_b \ket{sb}\bra{sb}$ which projects
onto the subsystem state $s$ in the parameter regime where the
subsystem is thermalised.  This can be explained theoretically (section
\ref{sec:ETHscaling}) using known results for the variance of the
overlap distribution.  This observation for $P_s$ then leads directly
to a thermalised reduced density matrix. (See section
\ref{sec:thermquant}.)

In this paper, we also highlight the importance to thermalisation of
the strength of interaction within the bath. We find that, at least
for our small bath and subsystem, a finite interaction strength is
needed for thermalisation. This is consistent with the expectation
that thermalisation is aided by inelastic scattering in the bath.

We point out that, although we have focussed on the thermalisation of
a spatially local subsystem, 
one can also study the thermalisation of few-body
observables over the entire system. This has been studied particularly in the
context of quantum quenches in a variety of
systems~\cite{Rigol2008,Rigol2010,Ji2011,*Fine2009,Canovi2011,Lesanovsky2010,*Ates2012},
including integrable
systems~\cite{Kinoshita2006,Rigol2007,Eckstein2008,Sotiriadis2009,Calabrese2011,Cassidy2011,Poilblanc2011}.
Moreover, one can discuss the dynamics of the relaxation towards a
thermal
state~\cite{Jensen1985,Eckstein2009,Genway2010,Lesanovsky2010,Ates2012,Kollar2011}. Both
of these issues are beyond the scope of this paper.

This paper proceeds as follows.  The following section introduces the
Hubbard model we study, discusses how the system is prepared initially
and provides a framework for studying thermalisation.  In
Section~\ref{sec:numerics}, we present a comprehensive set of results
for the thermalisation of two-site subsystems in small Hubbard
rings. We consider the effects of subsystem-bath coupling strength on
thermalisation and link these results to eigenstate thermalisation.
We further demonstrate the role of interactions between fermions
before exploring system size dependence and, finally, the energy width
of the initial prepared state.  Section~\ref{sec:eth} introduces
results from random matrix theory concerning the nature of the
eigenstates of the coupled system.  From these, we review the
arguments leading to eigenstate thermalisation~\cite{Srednicki1994}
and derive the scaling behaviour associated with the closeness to
perfect eigenstate thermalisation. In Section~\ref{sec:sizescaling},
we present an account of system-size scaling by considering a
threshold for non-perturbative mixing of uncoupled composite
eigenstates for the cases of both interacting and non-interacting
fermions. In Section~\ref{sec:expt}, we discuss implications for
experiments. Finally, in Section~\ref{sec:conclusions}, we give our
conclusions.

\section{The Model and Theoretical Framework}

We will consider closed quantum systems with unitary time evolution.
The system is prepared in an initial state $\ket{\Psi(t=0)}$ which
evolves in time $\ket{\Psi(t)} = e^{-iHt}\ket{\Psi(0)}$ under the
influence of the Hamiltonian $H$.
In this section, we discuss the specific model studied in this work
and our choice of initial states. Since we will investigate
thermalisation of a subsystem of this closed system, 
we will also discuss our criteria for a thermal state.  

\subsection{Hubbard Hamiltonian}
\label{sec:hamiltonian}

We divide the system into a local subsystem ($S$) and a bath ($B$).
The subsystem (bath) is described by a Hamiltonian ${H}_S$ (${H}_B$)
acting on the subsystem (bath) Hilbert space.  Let us denote the
subsystem (bath) eigenstates as $\ket{s}_S$ ($\ket{b}_B$) with
energies $\egysys_s$ ($\egybath_b$).  The subsystem and bath are
coupled by a Hamiltonian $\lambda V$.  We will use $\lambda$ as a
tunable parameter to control the strength of this coupling. 
At $\lambda=0$, the eigenstates are products of subsystem and bath
eigenstates, $\ket{sb}$, with energies $E_{sb} = \egysys_s +
\egybath_b$.  At non-zero $\lambda$, the eigenstates are in general
entangled with respect to the subsystem-bath partition. We denote
these composite eigenstates by $\ket{A}$ (using an uppercase index)
and their energies by $E_A$.

\begin{figure}[hbt]
\includegraphics[scale=0.2]{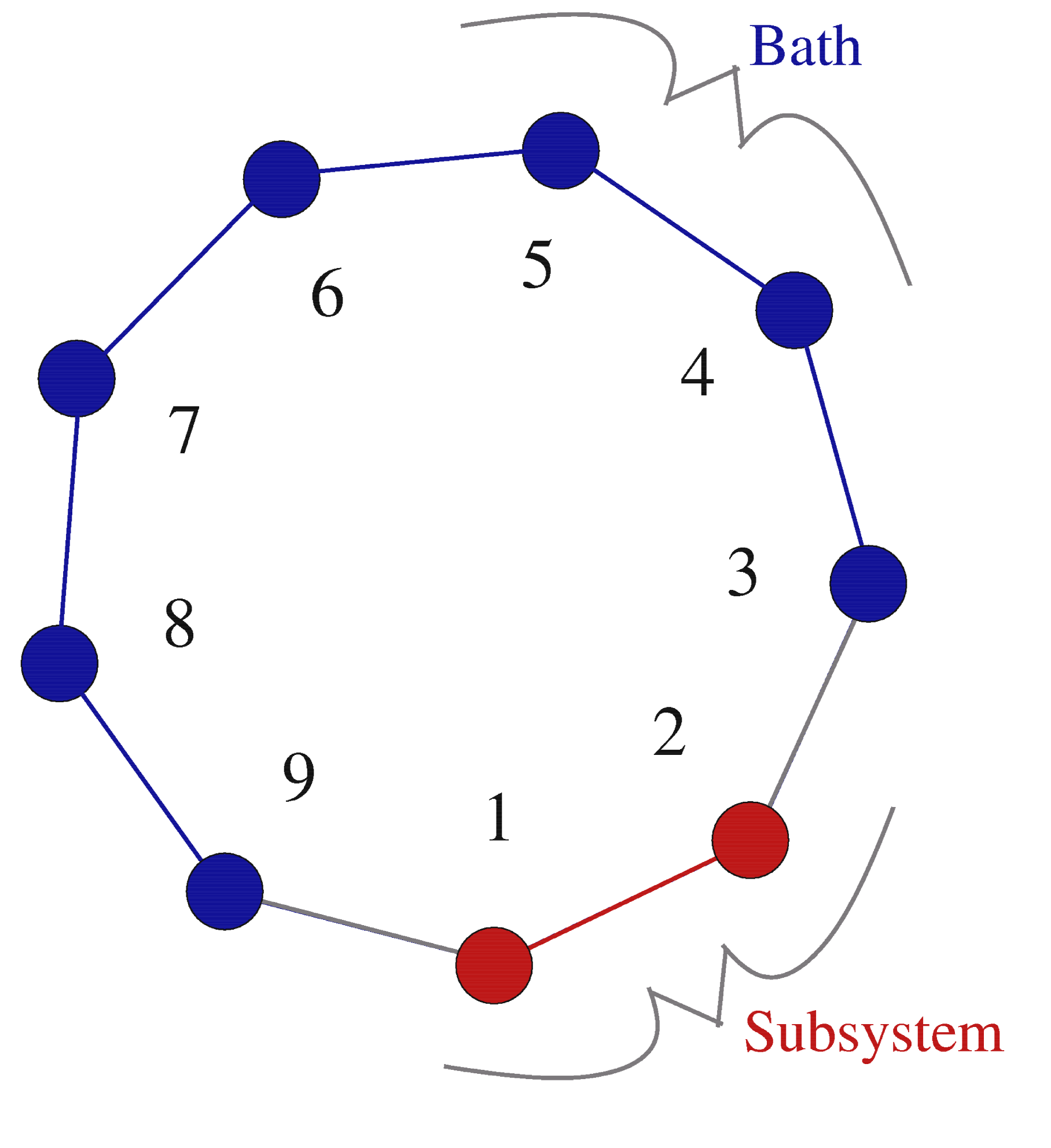}
\caption{Schematic diagram of a two-site subsystem in a lattice with 9
  sites ($L=9$).}
\label{lattice}
\end{figure}

In this work, we focus on the Hubbard model away from half filling 
as a simple model of interacting fermions. 
More specifically, we consist of a two-site subsystem in an
$L$-site Hubbard ring of fermions such that the Hamiltonian takes the
form $H = H_{S} + H_{B} + \lambda V$ with
\begin{align}
 H_S &=
 - \sum_{\sigma=\uparrow,\downarrow} 
 J_{\sigma} (c^{\dagger}_{1 \sigma} c_{2 \sigma} + \text{h.c.}) 
 +  U( n_{1 \uparrow} n_{1 \downarrow} +  n_{2 \uparrow} n_{2 \downarrow} ) \, , 
\notag\\
H_B &= - \sum_{i=3}^{L-1}
\sum_{\sigma=\uparrow,\downarrow} J_{\sigma} (c^{\dagger}_{i \sigma}
c_{i+1, \sigma} 
+ \text{h.c.}) +  U \sum_{i=3}^L n_{i \uparrow} n_{i \downarrow} \, , 
\notag\\
\lambda V &= -\lambda\sum_{\sigma=\uparrow,\downarrow} J_{\sigma}\left[ (c_{2 \sigma}^{\dagger} c_{3 \sigma} + c_{1 \sigma}^{\dagger} c_{L\sigma}) + \text{h.c.}\right]\,.
\label{eq:2linkH}
\end{align}
where $c^{\dagger}_{i \sigma}$ is a creation operator for a fermion
with spin $\sigma$ at
site $i$ and $n_{i\sigma} = c^{\dagger}_{i \sigma} c_{i \sigma}$ is the
number operator on site $i$ with spin $\sigma$. 
This Hamiltonian
describes a ring with the subsystem sites $i=1,2$ and bath sites $i=3$
to $L$ with two links between the subsystem and the bath.
Note that, in the case of $\lambda=1$, the Hamiltonian describes an
homogeneous ring.  We choose the hopping integrals $J_{\sigma}= J(1+\xi\;
\text{sgn}(\sigma))$, 
with $\xi=0.05$ to remove level degeneracies
associated with spin rotation symmetry. (We will use $J$ as the unit of
energy.) Breaking spin symmetry and the presence, in general, of
modified hopping integrals between sites $i=2$ and $3$, as well as
between sites $i=L$ and $1$ make this system non-integrable for non-zero
$U$.  
 
\begin{figure}[hbt]
\includegraphics[scale=1.0]{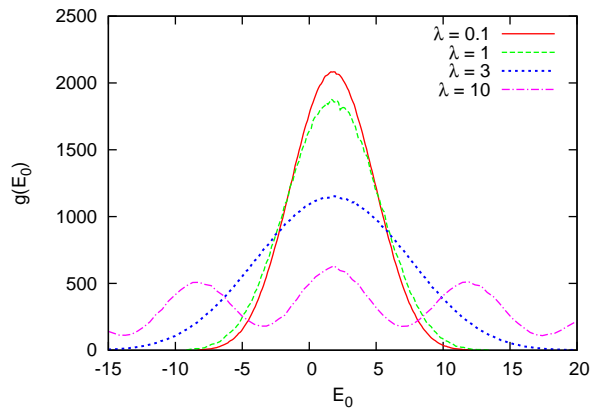}
\caption{The density of states $g(E_0)$ of the system at
  composite energy $E_0$ for different coupling strengths, $\lambda$
  (as labelled), for an $L=9$ site lattice where $U=J=1$. $g(E_0)$ is 
  generated as a histogram by counting eigenstates in a Gaussian window centered
  on $E_0$ with width $0.5J$.}
\label{DOS}
\end{figure}
\begin{figure}[hbt]
\includegraphics[scale=1.0]{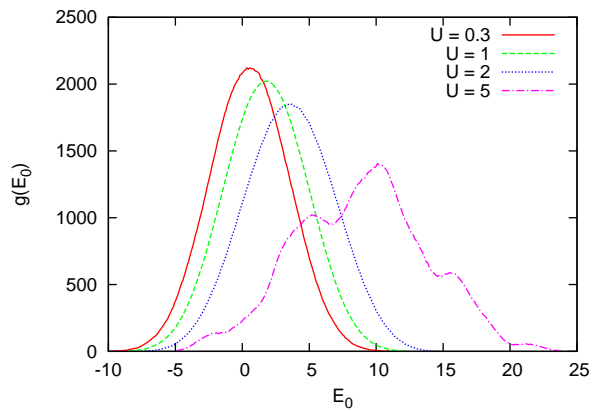}
\caption{The density of composite states $g(E_0)$ of the system at
  composite energy $E_0$ for different Hubbard interaction strengths,
  $U$ (as labelled), for an $L=9$ site lattice where $\lambda = 0.5$.
  $g(E_0)$ is generated as a histogram by counting eigenstates 
  in a Gaussian window centred on $E_0$ with width $0.5$. ($J=1$.)}
\label{DOS_U}
\end{figure}
The total particle number, $\Nparticle$, and spin component, $S^z$,
are conserved in addition to the total energy of the composite system.
In the numerical results we present, we consider lattices with up to
$L=9$ sites and with 8 fermions of total spin $S^z=0$. The two-site
subsystem has $\nstate_S=16$ eigenstates and the
7-site bath has 8281 eigenstates, while the composite 9-site system has a
total of $\nstate=15876$ states and an average level spacing $\Delta \simeq
10^{-3}J$. 

The spectrum of the composite system has
a smooth quasi-continuous density of states $g(E_0)$ for a range of
$\lambda$ and $U$.  This is illustrated in Figs.~\ref{DOS}
and~\ref{DOS_U}. 
The centre of the spectrum is located at $E_0 \simeq
1.77J$. The spectrum develops peaks for large $\lambda$ and $U/J$.
In the case of $\lambda \gg 1$, we attribute this to 
single-particle states with a large energy splitting proportional 
$\lambda J$ on the two links connecting the subsystem and bath. 
For $U/J\gg 1$, we attribute the peaks to a large energy gap to
doubly-occupied sites (sometimes referred to an `upper Hubbard band'
in the theory of strongly electron systems 
or doublons in the cold atoms literature).

\subsection{Initial States}
\label{sec:initialstates}

Throughout this work, we consider the composite system to be 
prepared in a pure state which is a product state of a subsystem state
and a bath state:
\begin{equation}
\label{eq:windowstate}
\ket{\Psi(t=0),E_0} =  
\ket{\phi}_S \otimes \sum_{b_i = b_l}^{b_u} \frac{1}{\sqrt{B}} \ket{b_i}_B\,.
\end{equation}
The initial subsystem state, $\ket{\phi}_S$, can be, for example,
$\ket{\!\uparrow,\downarrow}_S$ which is prepared with antiparallel
spins on sites $i=1$ and 2 of the lattice. The initial bath state
contains a linear combination of $B$ bath eigenstates $\ket{b_i}_B$.
These bath states are chosen to be within an energy window of bath
states such that  $\bra{\Psi}H\ket{\Psi} = E_0$.
The energy window has a width in energy of $\delta_B$ (specified by the state
indices in the range $b_l < b < b_u$). Unless stated otherwise, we
choose $\delta_B=0.5J$ which is small on the scale of
variations in the density of states. For a 7-site bath, this window
contains about 100 bath eigenstates.

As the system evolves in time, the state of the subsystem can be described
by the reduced density matrix (RDM) as defined in equation \eqref{eq:rdmdef}.
%
%
We will study the
reduced density matrix elements using the subsystem eigenstates
$\ket{s}_S$ as the basis: $\rdm_{ss'} = {}_S\bra{s}\rdm\ket{s'}_S$. 
We obtain the wavefunction $\ket{\Psi(t)}$ of the composite system 
using the eigenstates and energy eigenvalues from the
exact diagonalisation of the Hamiltonian $H$:
\begin{equation}
  \ket{\Psi(t)} = \sum_A e^{-iE_A t}\ket{A}\inner{A}{\Psi(0)}\,.
\label{eq:evolution}
\end{equation}

\subsection{Thermalisation}
\label{sec:thermdef}

To assess whether the subsystem reaches a thermal state, we must be
more precise about the criteria for a thermal state. We start with the
conventional definition of thermal equilibrium in the canonical
ensemble.  To define this `canonical thermal state', $\omega$, of the
subsystem, we set up the composite system at a total energy $E_0$ in a
microcanonical mixed state and consider the regime where the coupling
between the subsystem and the bath is negligible. In this way, the
thermal state of the subsystem may be determined by counting bath
states in an energy window, conserving the total energy and the global
$S^z$ and particle number $\Nparticle$. The reduced density matrix is diagonal
and is given by $\omega_{ss} = {}_S\bra{s}\omega\ket{s}_S$:
\begin{equation}
  \omega_{ss}=
  \frac{\nstate_b(E_0-\egysys_s,\Nparticle - \nparticle_s, S^z -
    s^z_s)}{\sum_{s'} 
    \nstate_b(E_0-\egysys_{s'},\Nparticle - \nparticle_{s'}, S^z - s^z_{s'})}
\label{eq:canonstate}
\end{equation}
where $\nstate_b(\egybath_b,\nparticle_b,s^z_b)$ is the number of bath
states with $\nparticle_b$ fermions of spin $s^z_b$ in a window of
width $\sim\delta_B$ centred on energy $\egybath_b$. This thermal RDM
is a function of $\egybath_s$, $\nparticle_s$ and $s^z_s$, arising
from the global conservation laws of the system.  For fixed $s^z_s$
and $\nparticle_s$ and for a range of subsystem energies $\egybath_s$
small compared with features in the density of states, the smooth density
of states allows us to write the RDM in the Boltzmann form
$\omega_{ss} \sim e^{-\egysys_s/T}$, with the inverse temperature
given by
\begin{equation}
\frac{1}{T} = \left.\frac{\partial \log \nstate_B(\egybath_b, \Nparticle-n_s,
    S^z - s^z_s)}{\partial \egybath_b}\right\vert_{\egybath_b = E_0}
\label{eq:temperature}
\end{equation}
We can in principle deduce a chemical potential and Zeeman field by
considering variations in $\nparticle_s$ and $s^z_s$. However, we are
considering small systems where the discreteness of these quantities
cannot be ignored and $\nstate_B$ is not a smooth distribution of $s^z_s$
and $\nparticle_s$. Nevertheless, Eq.~\eqref{eq:canonstate} may be used to
specify a thermal state of the subsystem for, in principle, any bath
size.

We note that such a Gibbs-like distribution has just three parameters,
differing from the `generalised Gibbs distribution' for integrable
systems~\cite{Rigol2007, Rigol2008, Cassidy2011}, where the number of
parameters extends with system size.

Let us now turn to the RDM that we obtain from the unitary evolution
from an initial pure state. 
We will be examining the behaviour of the RDM at long times.
It is useful to define the time average:
\begin{equation}
r_{ss} = \lim_{t\longrightarrow\infty} 
\frac{1}{t} \int_0^t dt' \,\, \rho(t')
\label{eq:rdef}
\end{equation}
If the reduced density matrix
reaches a steady state at long times, this state will be equal to the
time average $r$. We expect this to become diagonal.  Using
Eq.~\eqref{eq:rdmdef}, we see that
the diagonal
elements of the RDM are given by
\begin{equation}\label{eq:rhodecompose}
\bra{s}\rho\ket{s} = \sum_{ABb} e^{-i(E_A-E_B)t }
\inner{sb}{A}\inner{A}{\Psi(0)}\inner{\Psi(0)}{B}\inner{B}{sb}\,.
\end{equation}
Averaging over time for long times identifies $E_A$ with $E_B$.
Since we have lifted all symmetry-related degeneracies, this also
identifies states $A$ and $B$ in the sum above (barring accidental
degeneracies).
So we see that
\begin{align}\label{eq:r_eev}
r_{ss}&= \sum_A |\inner{\Psi(0)}{A}|^2 \bra{A}P_s\ket{A}\notag\\
\mbox{with\ }P_s &= \sum_b \ket{sb}\bra{sb}\,.
\end{align}
The operator $P_s$ projects from the composite
Hilbert space on to the subsystem state $\ket{s}_S$ by tracing over
bath states. 

The coefficients $\inner{\Psi(0)}{A}$ contain the information about
the initial state. However, this steady state may still be very close
to a state which is independent of initial conditions.  A sufficient
ondition is given by the `eigenstate thermalisation hypothesis'
\cite{Rigol2008}. Recall that from Eq.~\eqref{eq:windowstate} our
initial state $\ket{\Psi(0)}$ has been set up within a narrow energy
window. So, the overlap of $\inner{\Psi(0)}{A}$ should be only
non-zero in a window of eigenenergies. (We will give a more
quantitative discussion of the width of this window in Section
\ref{sec:ldos}.)  The eigenstate thermalisation hypothesis assumes
that $\bra{A}P_s\ket{A}$ for a system with composite energy $E_0$
depends only weakly on the choice of $\ket{A}$ in this window of
eigenenergies.  This allows us to replace $\bra{A}P_s\ket{A}$ by its
average value $\overline{\bra{A}P_s\ket{A}}$ over the eigenenergy
window.  In that case, $r_{ss} \simeq \overline{\bra{A}P_s\ket{A}}
\sum_A |\inner{\Psi(0)}{A}|^2 = \overline{\bra{A}P_s\ket{A}}$. Thus,
we see that the steady state $r_{ss}$ may indeed be independent of
initial conditions.

Furthermore, if $\overline{\bra{A}P_s\ket{A}}$ is close to the value
$\omega_{ss}(E_0)$ for the canonical ensemble \eqref{eq:canonstate},
then Eq.~\eqref{eq:r_eev} can be written as $r_{ss} \simeq
\omega_{ss}(E_0)$.  Thus, $r_{ss}$ will be close to the canonical
thermal state $\omega$ for any initial state with a definite energy.

In summary, we break down the question of whether the subsystem
thermalises into four criteria similar to the ones in
Ref.~\cite{Linden2009}.  Our criteria are
\begin{enumerate}
\item Firstly, we should establish that the reduced density matrix 
  reaches a steady state at long times.
\item The steady state should be diagonal in the subsystem energy eigenbasis
  with all off-diagonal elements falling to zero for long times. This
  demonstrates a loss of quantum coherence. 
\item This steady state should have no memory of the initial state, 
  such as the precise way in which the subsystem or the bath is prepared.
\item Finally, we ask if this steady state is close to the canonical
  thermal state $\omega$. If this is the case, we will say the system
  exhibits `canonical thermalisation'.
\end{enumerate}

We are leaving open the possibility
that the subsystem reaches a steady state with no memory of the
initial state, but does not resemble the canonical thermal state. This
may be possible since the canonical state has been derived assuming
the bath states are unperturbed by the coupling with the subsystem
which may not hold in the small quantum systems studied here when the
coupling $\lambda$ is of order unity.

We explore these questions with numerical studies of the Hubbard
Hamiltonian in the following section. We will discuss the more
sophisticated picture of the eigenstate thermalisation hypothesis
separately in Section~\ref{sec:eth}.

\section{Numerical Results}
\label{sec:numerics}

\subsection{Long time behaviour}
\label{sec:longtime}

\begin{figure}[hbt]
\includegraphics[scale=0.8]{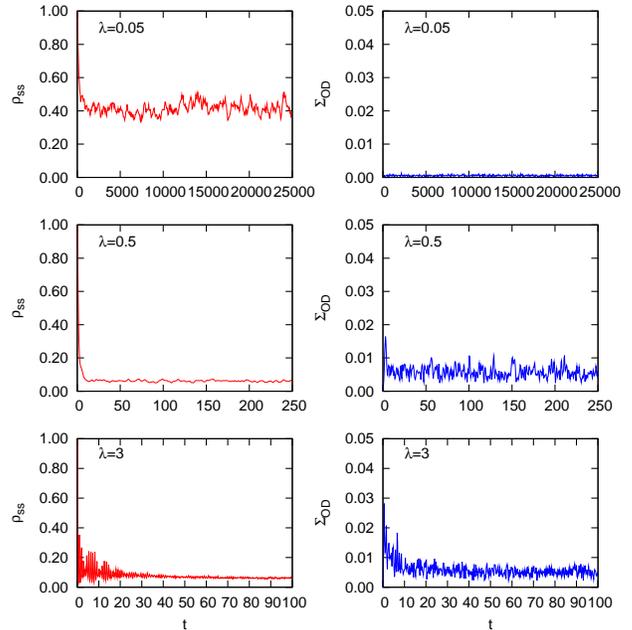}
\caption{Left: Time dependence of the initial state occupation probability
  $\rdm_{ss}$ for initial state $\ket{s}_S = \ket{\uparrow,\uparrow}$
  for three coupling strengths $\lambda=0.05$, 0.5 and 0.3.  
  Right: a measure of the magnitude of the
  off-diagonal elements $\Sigma_{OD}$, defined by
  Eq.~\eqref{eq:SigmaOD}. 
  Interaction strength $U=J$, width of the initial bath state 
  $\delta_B=0.5J$, total system energy $E_0=-2J$, size $L=9$.}
\label{timeplot}
\end{figure}
We proceed to demonstrate that the first two criteria for
thermalisation listed at the end of Section~\ref{sec:thermdef}) 
are met for a range of
system parameters. These requirements are that the reduced density matrix
$\rdm$ should approach a steady state at long times, with 
its off-diagonal elements falling to zero.

Initial states of the form in Eq.~\eqref{eq:windowstate} were
constructed with the initial subsystem state
$\ket{\uparrow,\uparrow}_S$.  It was found that evolving $\rdm(t)$ in
time under the Hamiltonian $H$ results in almost steady states for a
wide range of $\lambda$, provided the composite energy is not close to
the edge of the spectrum. With interaction strength $U=J$ and bath
width $\delta_B=0.5J$, this range is $0.05 \apprle \lambda \apprle
3$ for a composite energy $E_0 = -2J$.  Examples are shown in the left
panel in Fig.~\ref{timeplot}. This shows the relaxation of the
diagonal element of $\rdm$ corresponding to the initial-state
occupation probability for couplings $\lambda =0.05$, 0.5 and 3. The
dynamics of the relaxation is fast and featureless for $\lambda$ up to
1. The temporal fluctuations around the long-time steady state are small at
this energy $E_0 = -2J$.  In fact, if we use an energy close to the
centre of the spectrum of the composite system ($E_0=1.77J$), temporal
fluctuations are even smaller for a given $\lambda$.  On the other
hand, for an energy closer to the edges of the spectrum, the density
of states is small so that few composite states construct the initial
state. The presence of only a few frequencies in the time evolution
limits the closeness to a steady state achievable.

We see beating oscillations at $\lambda=3$ which we attribute to
the strongly split single-particle states at the two subsystem-bath
links at large $\lambda$. Indeed, for even larger $\lambda$, $\rho(t)$
no longer reaches a steady state, and the frequency spectrum begins to
show peaks at frequencies which are integer multiples of $\lambda
J$.

Let us now investigate the second condition for thermalisation that
off-diagonal elements fall to zero with only small temporal fluctuations, as
predicted in~\cite{Gogolin2010}. We compute the root-mean-square sum of
these off-diagonal elements:
\begin{equation}
\Sigma_{OD}(t) = \sqrt{\sum_{s<s'} |\rdm_{ss'}(t)|^2}\,.
\label{eq:SigmaOD}
\end{equation}
This is shown in the right panel of Fig.~\ref{timeplot}.
By construction, $\Sigma_{OD}$ is larger than any single off-diagonal
element. We see that the effect of off-diagonal
elements can be neglected at long times: $\Sigma_{OD}$ is, with decreasing
$\lambda$, shown to be from $\sim 10^{-1}$ down to $\sim 10^{-3}$
times smaller than each diagonal element. 

Having established that, within a range of coupling strengths, 
the subsystem RDM does reach a diagonal steady state
with only small temporal fluctuations, we will now use
Eq.~\eqref{eq:r_eev} to compute the steady-state form $r$ without
explicitly computing $\rho(t)$ at many times and taking a time
average. This is less computationally expensive and provides
a definitive long-time subsystem state without the need for
numerically averaging out small temporal fluctuations.

We will now proceed to 
explore two further requirements of thermalisation: these are the
extent of initial-state independence and closeness to the thermal
state $\omega$.  The effective temperature of the subsystem may also
be estimated from $r$.

\subsection{Quantifying Thermalisation}
\label{sec:thermquant}

Next we develop measures to characterise the extent to which the third
and fourth of our thermalisation criteria, listed in
Section~\ref{sec:thermdef}, are met. 

Criterion 3 in Section~\ref{sec:thermdef} is concerned with the loss
of memory of the initial state at long times. To quantify the
variation in the steady state due to different initial states, we
introduce the measure $\Delta r$ which measures the root-mean-square
variation in diagonal reduced density matrix elements for different
initial subsystem states:
\begin{equation}
\label{eq:Deltar}
\Delta r = \frac{1}{2} 
\sum_s [{\expectation{{r_{ss}}^2} - \expectation{{r_{ss}}}^2}]^{\frac{1}{2}}
\end{equation}
with $\expectation{\ldots}$ denoting an average over all $16$ initial states 
in the subsystem Fock basis, as in Eq.~\eqref{eq:sigmaomega}. We expect $\Delta r$ to be small
when the long-time steady state no longer depends on how the system was initially prepared.

Criterion 4 in Section~\ref{sec:thermdef} addresses the closeness of the
subsystem state at long times to the canonical thermal state $\omega$. We
quantify this with the
quantity $\sigma_{\omega}$, defined as:
\begin{equation}
\label{eq:sigmaomega}
\sigma_{\omega} = \frac{1}{2} \sum_s 
\expectation{\left| {r_{ss}} - \omega_{ss} \right|}\,.
\end{equation}
Here, $\expectation{\ldots}$ denotes an average over all $16$ initial
states in the subsystem Fock basis (eigenstates at $J=0$).  As such,
this is a measure of the average distance to the thermal state,
$\omega$, for the set of initial subsystem states spins localised on
the lattice sites.  It is a special case of a more general distance
measure~\cite{Linden2009}, $\langle\frac{1}{2} \text{Tr} \sqrt{( {r} -
  \omega )^2}\rangle$, which equals $\sigma_{\omega}$ in the case
where the elements of $r$ in the subsystem eigenbasis form diagonal
matrices.  As established above, this is the case for $0.05\apprle
\lambda \apprle 3$.  Within this range, we may interpret
$\sigma_{\omega}$ as the probability, upon making measurements on the
subsystem, that $r_{ss}$ could be distinguished from
$\omega_{ss}$~\cite{Popescu2006}.

From the definitions of these two measures, it is clear that if
$\Delta r$ is large then $\sigma_{\omega}$ is necessarily large too:
if there is a large variation in $r$ for different initial states,
many of these states must be far from the uniquely defined 
canonical thermal state $\omega$.
Conversely, it is possible for $\sigma_{\omega}$ to be large with
$\Delta r$ small, because the subsystem may relax consistently to a
state $r$ other than the canonical state $\omega$.

We will also compute the von Neumann entropy of the subsystem.
Because off-diagonal elements of $\rho(t)$ are
virtually zero at long times even for very small $\lambda$, 
we introduce an initial-state-averaged
subsystem entropy for the equilibrium state, which we define by
\begin{equation}
S = - \sum_s \expectation{ r_{ss} \log r_{ss} }
\end{equation}
where $\expectation{\ldots}$ denotes an average over all initial
subsystem states in the subsystem Fock basis.

We would also like to characterise subsystems showing thermalisation
with an effective temperature. As discussed in
Section~\ref{sec:thermdef}, if we consider the subsystem at a given
particle number $\nparticle_s$ and spin $s^z_s$, we expect the
steady-state RDM, $r$, to approach the Boltzmann form
\eqref{eq:temperature} for $\omega$ if the subsystem relaxes to the
canonical thermal state $\omega$. Therefore, we extract an effective
temperature $\Teff$ from the RDM, $r$, of the steady states that we
find using a least-squares fit to the form
\begin{equation}
\log r_{ss} = -\frac{\egysys_s}{\Teff} +\text{const}\,.
\label{eq:tempfit}
\end{equation}
We will focus on the four-state subsector with $\nparticle_s=2$ and $s^z_s =
0$ because it is the subsector with the largest number of bath
states. Note that it is possible that we can have a good fit to this
form with an effective temperature even if the steady state is not
close to the canonical state $\omega$.

\subsection{The Role of Coupling Strength}
\label{sec:couplingstrength}

\begin{figure}[!htb]
 \begin{centering}
   \includegraphics[scale=1.35]{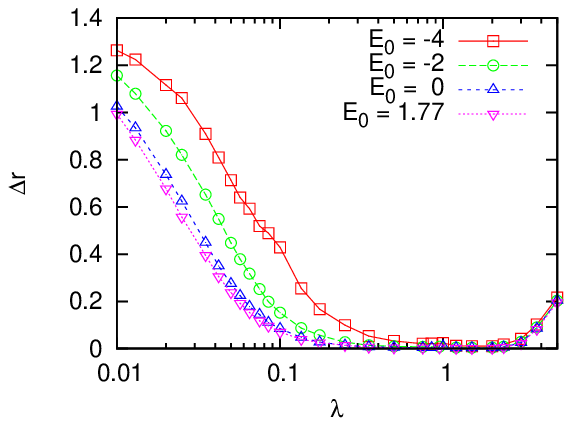}
   \includegraphics[scale=1.35]{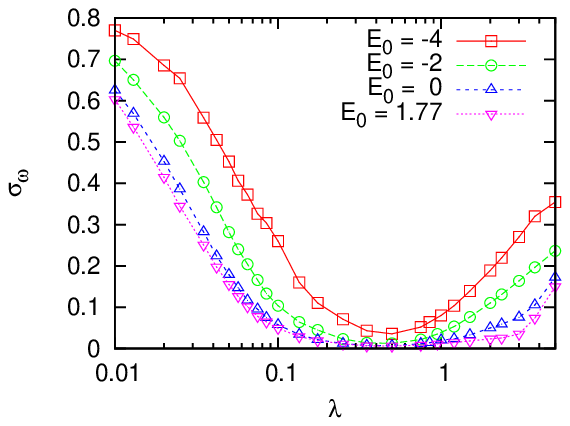}
   \includegraphics[scale=1.35]{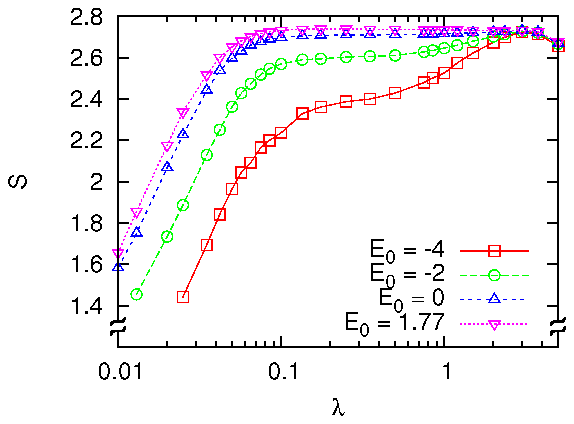}
   \caption[$\sigma_{\omega}$, $\Delta r$ and $S$]{
     Memory of initial state $\Delta r$, closeness to the thermal
     state $\sigma_{\omega}$ and the subsystem entropy $S$
     as a function of coupling strength $\lambda$ for different 
     composite energies $E_0$. ($U=J=1$, $\delta_B=0.5$, $L=9$.)}
   \label{thermvars}
 \end{centering}
\end{figure}

In the previous section, we discussed how we measure the memory of the
initial conditions ($\Delta r$) in the steady state, closeness
($\sigma_\omega$) to the canonical thermal state, the effective
subsystem temperature ($\Teff$) and the entropy ($S$) of the
subsystem. We will now discuss how these measures of thermalisation
change over a broad range of subsystem-bath coupling strengths
$\lambda$. We show results at different total energies $E_0$ between
$-4J$ and 1.77$J$.

In Fig.~\ref{thermvars}, we present our results for $\Delta r$,
$\sigma_{\omega}$ and $S$ as a function of the coupling strength
$\lambda$ (for a system with $U=J$ and an initial state of bath width
$\delta_B = 0.5J$).  Our results for $\Delta r$ demonstrate that the
subsystem reaches a steady state with little dependence on initial
conditions over a wide range of coupling strengths $\lambda$. We see
significant dependence on initial state beyond this range, at both
small and large $\lambda$.  Moreover, our results for
$\sigma_{\omega}$ show that the subsystem reaches the canonical state
$\omega$ over a similar, albeit slightly narrower, 
range of coupling strengths.  Outside this
range, the long-time steady state shows strong deviation from the
canonical state.

\begin{figure}[!htb]
  \begin{centering}
    \includegraphics[scale=0.4]{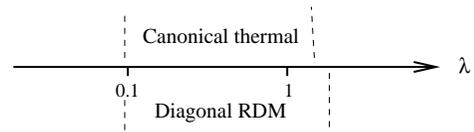}
    \caption{A schematic diagram indicating the range in
      $\lambda$ where the subsystem reduced density matrix, $r$,
      is diagonal, and where it is close to the canonical
      thermal reduced density matrix, $\omega$.}
    \label{phasediagram}
  \end{centering}
\end{figure}
In the coupling range where $\Delta r$ and $\sigma_\omega$ are both
small, we find that the entropy $S$ reaches a plateau as a function of
$\lambda$.  Beyond this range at low $\lambda$, the subsystem entropy
$S$ drops with decreasing $\lambda$. This is consistent with the
subsystem retaining information of its initial conditions. On the other
hand, the entropy rises when $\lambda$ is increased beyond the
plateau.  The asymmetry between low and high coupling indicates that
the departure from thermalisation at small and large $\lambda$ have
different physical origins, as we discuss later.
The behaviour of the subsystem reduced density matrix  as a function
of $\lambda$ is summarised by the schematic diagram in
Fig.~\ref{phasediagram} for the range
of energies shown in Fig.~\ref{thermvars}.

\begin{figure}[!htb]
 \begin{centering}
   \includegraphics[scale=1.05]{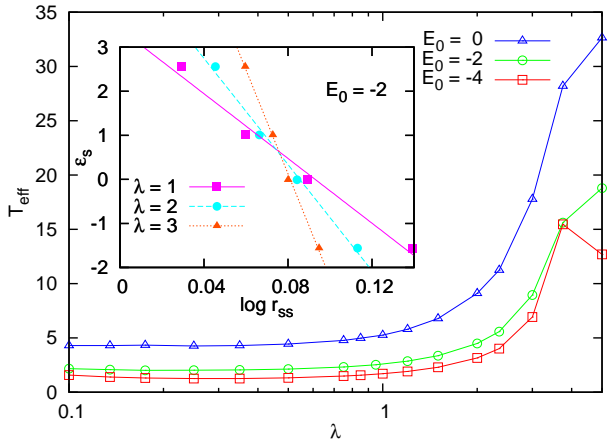}
   \caption[Effective temperature]{
     Effective temperature $\Teff$ as a function of
     coupling strength $\lambda$ in the thermalised regime
     for different composite energies
     $E_0$. Inset: example of fit of $r_{ss}$ to the Boltzmann form \eqref{eq:tempfit}. ($U=J=1$, $\delta_B=0.5$, $L=9$.)}
   \label{Teff}
 \end{centering}
\end{figure}

We show in Fig.~\ref{Teff} the effective temperature $\Teff$ extracted
at different energies $E_0$ using the fit in
Eq.~(\ref{eq:tempfit}). We include only the range of coupling
strengths where the fit is reasonable.
It is noteworthy that our results with
two lowest energies, $E_0=-4J$ and $-2J$, show effective temperatures
close to the degeneracy temperature, approximately $2J$ for this
Hubbard system near half filling.

We have not shown $\Teff$ for the highest energy we used, $E_0 =
1.77$.  This energy corresponds to the centre of the energy spectrum
for all $\lambda$ plotted.  At this energy, all the states of the
subsystem have nearly equal statistical weight at this energy. In
other words, the effective temperature is nearly infinite.  This is also
reflected in the subsystem entropy (Fig.~\ref{thermvars}) which is
close to $\log 16$ at $E_0=1.77$, as expected for our 16-state
subsystem at high temperatures.

For systems exhibiting canonical thermalisation (small
$\sigma_\omega$), the effective temperature $\Teff$ is approximately
independent of the coupling strength up to $\lambda \simeq 1$. In
fact, this effective temperature is close to the canonical temperature
defined in Eq.~(\ref{eq:temperature}) by counting bath states
in the limit of $\lambda \to 0$, as reported in \cite{Genway2010}. As
already mentioned, we find in this regime that the subsystem entropy
(Fig.~\ref{thermvars}) is also roughly independent of $\lambda$.

We will now turn to the crossover from non-thermalisation to
thermalisation as we increase the coupling strength from zero.  We can
choose a rough measure of the threshold, $\lambdathermal$, for this
crossover as the coupling strength at which $\sigma_\omega$ drops
below 25\%. Alternatively, we can use the coupling
strength at which the subsystem entropy reaches a plateau in
Fig.~\ref{thermvars}.  At $E_0 = -2J$, we find $\lambdathermal \simeq
0.05$. At the lower energy $E_0=-4J$, $\lambdathermal$ is higher at
approximately 0.1. At the energy $E_0=1.77J$ corresponding to the centre of the
spectrum, $\lambdathermal$ is smallest at 0.03. 
The crossover between memory and lack of memory of the initial
state also occurs around this characteristic coupling
$\lambdathermal$. (We discuss this criterion further in
Section~\ref{sec:sizescaling}.)  That thermalisation does not occur
for small coupling strengths is because of the finite level spacing,
$\Delta$, in the finite-size bath.  Physical intuition might suggest
that subsystem-bath couplings, however weak, allow relaxation in
subsystems. This is a reasonable assertion for systems with
macroscopic baths where the bath spectrum is
quasi-continuous. However, for a small system with a non-zero level
spacing at weak coupling, the eigenstates of the composite system may
only be slightly perturbed from the decoupled subsystem-bath product
states $\ket{sb}$ if the typical matrix elements mixing these product
states are small: $\bra{sb} \lambda V \ket{s'b'} \ll \Delta$.  In this
weak-coupling limit, thermalisation cannot occur from an initial
subsystem state $\ket{\phi}_S$, when the composite eigenstates are all
close to product states of the form
$\ket{\phi}_S\otimes\ket{b}_B$. The system would retain strong memory
of the initial state. Therefore, we expect a non-zero threshold for
thermalisation for a finite system. We will examine more
quantitatively the overlap of the composite eigenstates with the
decoupled product states in Section~\ref{sec:eth} and we will compare
the empirical $\lambdathermal$ extracted here with a theoretical
estimate in Section~\ref{sec:ETHscaling}.

Let us now turn to the strong-coupling regime of $\lambda\gg 1$. As
already discussed in Section~\ref{sec:longtime}, the system does not
reach a steady state at very high $\lambda$, and so it is not
thermalised. We believe that this is a boundary effect in the sense
that the dynamics in our `subsystem' consisting of sites 1 and 2
become altered at very large $\lambda$ because of the very large
hopping on the links between sites 2 and 3 and between sites 1 and
$L$. As already discussed in Section~\ref{sec:hamiltonian},
single-particle states localised on these links become visible as a
feature the composite density of states at $\lambda=10$
(Fig.~\ref{DOS}). We believe that the four sites ($i=L$,1,2,3) will
thermalise as a cluster in the sense that it has a canonical reduced
density matrix, provided that the bath of size $L-4$ is sufficiently
large. Nevertheless, since the eigenstates of the two-site cluster and
the four-site cluster are very different at large $\lambda$, the
thermalisation of the four-site cluster does not imply a diagonal RDM
for the two-site cluster.  In any case, we wish to make the point that
this lack of thermalisation at large coupling is qualitatively
different in origin from the lack of thermalisation at small coupling.

It is interesting to examine more closely 
the departure from thermalisation as we
increase $\lambda$ in the range of $\lambda$ between 1 and 3
for $U=J$ (see $\sigma_\omega$ in Fig.~\ref{thermvars}). In this
crossover region, we find steady states that have lost memory of the
initial state (small $\Delta r$) but these states deviate from the
canonical thermal state $\omega$, as can be seen in a rising 
$\sigma_\omega$ as $\lambda$ is increased beyond unity. 
Moreover, the RDM has a reasonable fit to the Boltzmann
form \eqref{eq:tempfit}, although the fitted temperature departs
significantly from the canonical temperature \eqref{eq:temperature}.  One
can say that the system is still in an `effective' thermal state in this
crossover regime. We will return to this in Section
\ref{sec:eigentherm}.

\begin{figure}[!hptb]
 \begin{centering}
   \includegraphics[scale=1.35]{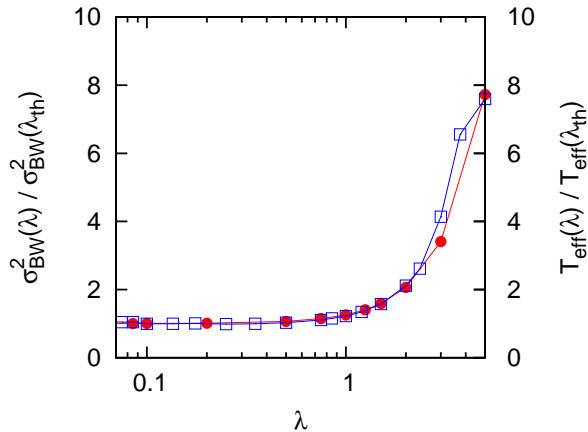}
   \caption{Gaussian width of the density of states, $\sigma_{BW}$ 
     (solid circles) 
     and effective temperature $\Teff$ (hollow squares)  
     as a function of coupling strength $\lambda$. 
     ($E_0=0$, $U=J=1$, $\delta_B=0.5$, $L=9$.) 
     Both quantities are normalised to their values at $\lambdathermal$.}
   \label{tempscaling}
 \end{centering}
\end{figure}

Interestingly, we observe that this crossover regime tracks closely a
decrease in the density of states of the composite system. The density
of states (Fig.~\ref{DOS}) can be approximated as a
Gaussian:
\begin{equation}
  g(E_0) \propto 
  \exp\left(-{\frac{(E_0 - \overline{E_0})^2}{2\sigma_{BW}^2}}\right)\,
\end{equation}
where $\overline{E_0}$ is the energy of the band centre, and
$\sigma_{BW}$ can be used as a measure of the width of the Gaussian.  We
see in Fig.~\ref{tempscaling} that $\sigma_{BW}$ rises sharply as we
increase $\lambda$ beyond unity, similar to the behaviour of the
fitted effective temperature (Fig.~\ref{thermvars}). In fact, $\Teff
\propto \sigma_{BW}^2(\lambda)$, as seen in Fig.~\ref{tempscaling}
where we show the two quantities normalised to their
($\lambda$-independent) values at small $\lambda$. Note that, at weak
coupling and at fixed energy $E_0$, the derivative
\begin{equation}
\frac{\partial \log g(E)}{\partial E}\bigg|_{E=E_0} 
= \frac{E_0 - \overline{E_0}}{\sigma^2_{BW}}
\end{equation}
can be associated with the inverse temperature of a bath of size
$L=9$. In other words,  
it appears that the effective temperature of the
subsystem is better described by the canonical temperature of the
whole system, instead of just the bath.
This result is not surprising in this regime 
where the coupling of our 2-site subsystem to
the $L=7$ chain is of order unity, since 
the distinction between subsystem and bath is blurred. 

\subsection{Dependence on Interaction Strength}
\label{sec:interactionstrength}

We now turn to the effects of the particle-particle interaction
strength, $U$, on thermalisation.  In the results which follow, the
coupling strength is fixed, as previously, at $\lambda = 0.5$ and we
will also fix the bath width at $\delta_B = 0.5J$.  The energy of the
composite system, $E_0$, will be fixed such that it is always at the
peak in the centre of the composite spectrum, at $E_0 \approx 2U$.
This is necessary since the shape of the spectrum is a strong function
of $U$ and the density of states at a given energy can vary
significantly.  The effects of the interaction strength on the
composite density of states are shown in Fig.~\ref{DOS_U}.  We find
that for $U\apprge 4J$, peaks separated by $U$ appear.  If comparisons
were to be made between different $U$ for initial states at fixed
$E_0$, the features in the density of states which evolve with $U$
would introduce unwanted artefacts. Even in the centre of the
spectrum, it should be noted that there is a fall in the density of
states at the central peak at $E_0 \approx 2U$, which occurs over a
range $1 \apprle U/J \apprle 4$ due to an overall broadening of the
density of states (see Fig.~\ref{DOS}).

\begin{figure}[!htb]
 \begin{centering}
   \includegraphics[scale=1.4]{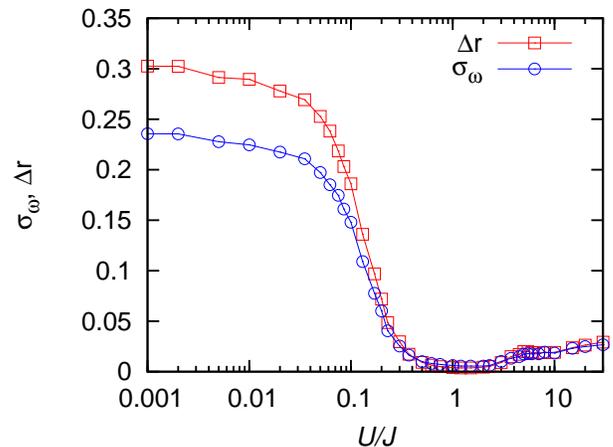}
   \caption{Memory of initial state, $\Delta r$ , and closeness to the
     canonical thermal state, $\sigma_{\omega}$, as a function of interaction $U/J$
     for $\lambda = 0.5$.  The composite energy $E_0$ is chosen to be
     fixed on the central maximum in $g(E_0)$, which lies close to
     $2U$.  $\delta_B=0.5J$.}
   \label{thermvars_U}
 \end{centering}
\end{figure}

To measure thermalisation at different $U$, we will again employ the
measures $\Delta r$  and $\sigma_{\omega}$ as defined in
Eqs.~\eqref{eq:Deltar} and \eqref{eq:sigmaomega}.  The effective
temperature is not shown since the initial-state energy is at a
maximum in the density of states which corresponds to infinite
subsystem effective temperature.  In Fig.~\ref{thermvars_U}, we
demonstrate that thermalisation, independent of the initial state, is
found for $U \apprge 0.1$.  There is a broad minimum in plots of both
$\Delta r$ and $\sigma_\omega$, defined by the lack of thermalisation
at small $U$ and a small increase in the plotted quantities over the
range $1\apprle U/J \apprle 5$.  

The slight increase in $\Delta r$ and $\sigma_\omega$ above $U\simeq
J$ coincides with the falling density of states in the centre of the
spectrum shown in Fig.~\ref{DOS}. So, the increase may be partly
associated with the reduction in the number of states in the fixed
bath window of our initial state.

The behaviour at small $U \apprle 0.1J$ cannot be similarly related to
the density of states.  However when $U = 0$, the nature of the
coupling is very different because the bath states are Slater
determinants single-particle states.  The single-particle level
spacing is large compared to $\lambda J$ if $\lambda\ll 1$. So, we
expect that thermalisation is poor for small non-interacting
systems. In other words, for small $U$, we need larger system sizes to
observe thermalisation. We will present our data for different system
sizes in the next subsection (Fig.~\ref{thermvars_U_sizes}).

\subsection{System Size Dependence}
\label{sec:thsc}

It is interesting to study thermalisation as a function of system
size. Owing to the exponential dependence of the Hilbert-space
dimension on lattice size, it is not possible to find the full
spectrum of large lattice. We will instead concentrate on the loss of
thermalisation as we reduce the system size.  If we use even smaller
systems, reducing the number of sites rapidly leads to Hilbert spaces
so small that thermalisation is not observed at all. Nevertheless, our
results show that thermalisation is possible in surprisingly small
systems, as long as $U\simeq J$ so that the system is not close to the
non-interacting limit, and as long as the density of states is not too
low.  To attempt to see some effects of reducing system size on
thermalisation, we consider composite states prepared with energies
$E_0$ in the centre of the band where the density of states is
highest.  As with our studies of the effects of interaction strength,
this also eliminates unwanted effects due to the changing bandwidth
with system size.

\begin{figure}[!htb]
 \begin{centering}
   \includegraphics[scale=1.4]{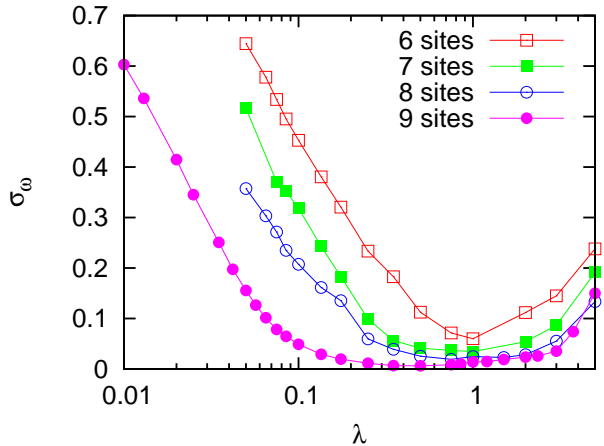}
   \caption{Closeness to canonical thermal state, $\sigma_{\omega}$, 
     as a function of coupling strength $\lambda$ for
     different system sizes $L$, for composite energies
     $E_0$ in the centre of the composite energy spectrum.  The number of
     particles was selected to keep $S^z=0$ with the number of particles 
     equal to $L$ and $L-1$ respectively for even and odd $L$.  
     ($U=J=1$, $\delta_B=0.5$.)}
   \label{thermsites}
 \end{centering}
\end{figure}

First of all, let us consider how our results in Section
\ref{sec:couplingstrength} for the dependence on coupling strength
changes with system size.  Shown in Fig.~\ref{thermsites} are plots of
$\sigma_{\omega}$ for different lattice sizes down to six sites.  In
each case, the subsystem size was fixed at two sites and the initial
bath width was fixed at $\delta_B = 0.5J$. Interestingly, for
$\lambda\simeq 1$, thermalisation is maintained down to a four-site
bath.  However, the range of couplings over which thermalisation
occurs is greatly reduced.  We return to system-size scaling in
section~\ref{sec:sizescaling}.

\begin{figure}[!htb]
 \begin{centering}
   \includegraphics[scale=1.4]{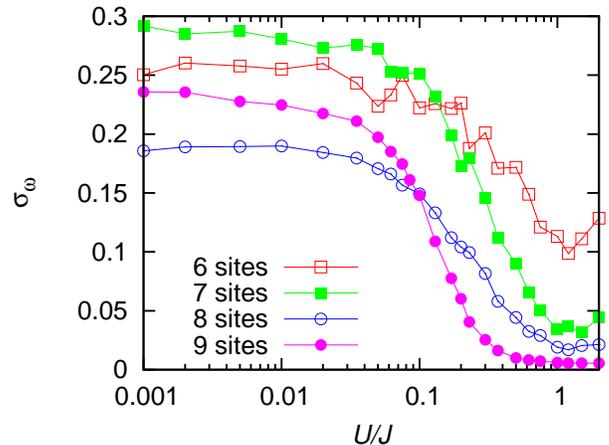}
   \caption[$\sigma_{\omega}$ vs $U$ for different sizes.]{
     Closeness to canonical thermal state, $\sigma_{\omega}$, as a function of
     interaction $U/J$ for different system sizes. $\lambda = 0.5$.}
   \label{thermvars_U_sizes}
 \end{centering}
\end{figure}

We can also see how our results in Section
\ref{sec:interactionstrength} for the dependence on interaction
strength, $U$, change with system size. We see in Fig.~\ref{thermvars_U_sizes}
that the larger systems
have a wider range of interaction strengths over which the system
approaches the canonical thermal state (small
$\sigma_\omega$). Moreover, $\sigma_\omega$ is lower for larger
systems at a given $U$. This is consistent with our expectation that
weakly-interacting systems require larger system sizes for
thermalisation.
We will explore system-size scaling in 
Section~\ref{sec:sizescaling}.

\subsection{Dependence on Initial Bath State}

\begin{figure}[!hbt]
 \begin{centering}
   \includegraphics[scale=1.2]{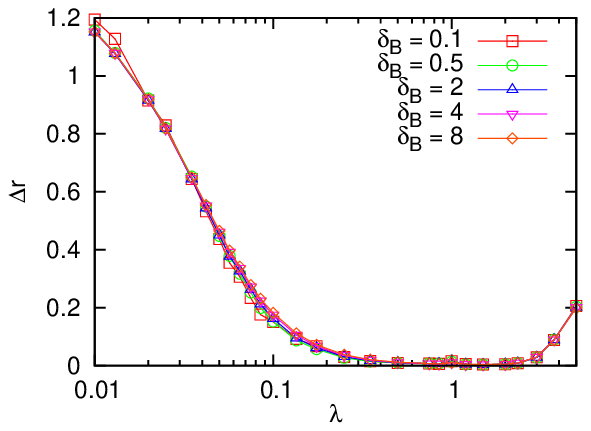}
   \includegraphics[scale=1.2]{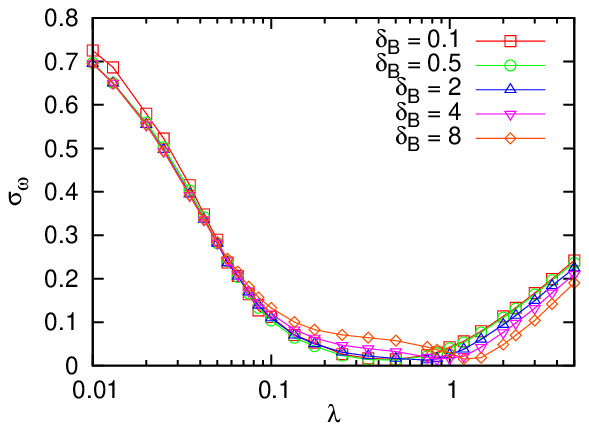}
   \includegraphics[scale=1.2]{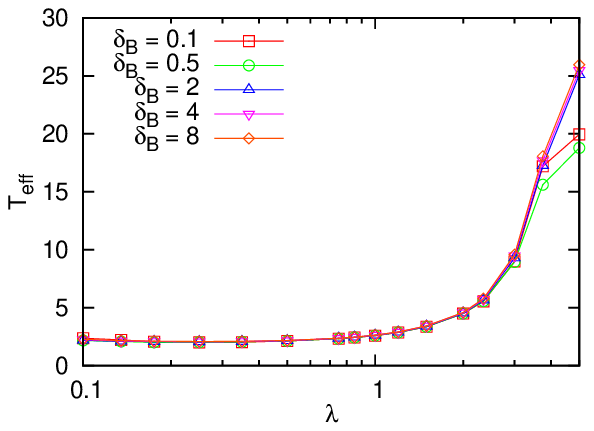}
   \caption[$\sigma_{\omega}$, $\Delta r$ and $\Teff$ for different
   bath widths $\delta_B$.]{$\Delta r$, $\sigma_{\omega}$ and $\Teff$
     as functions of coupling strength $\lambda$ for different
     bath-window widths $\delta_B$. The composite energy
     $E_0=-2J$. The average level spacing $\Delta/J \approx 10^{-3}$. ($J=1$)}
   \label{thermvars_W}
 \end{centering}
\end{figure}

The thermal state should not depend on the microscopic details of the
initial bath state.  We will now demonstrate that the
thermalisation behaviour found at long times is independent of the
initial bath state. More specifically, we will vary the energy width
$\delta_B$ of the initial bath state.  For all of the numerical
results presented thus far, we have considered initial states of the
form~\eqref{eq:windowstate} where the initial bath state is a pure
state, with components in the bath eigenbasis non-zero only in a
window of width $\delta_B=0.5J$.  This was chosen because it is small
compared with changes in the density of states. In
Fig.~\ref{thermvars_W}, we show plots of $\sigma_\omega$, $\Delta r$
and $\Teff$ against $\lambda$ for values of $\delta_B$ spanning almost
two orders of magnitude. In other words, these are results for vastly
different bath states with the only constraint that they should be
centered at the same energy.

We find that the thermalisation behaviour is essentially
$\delta_B$-independent for a broad range in $\delta_B$.  Remarkably
even up to $\delta_B=8J$, approximately half of the width of the
composite eigenspectrum, we see $\delta_B$ makes virtually no
difference to the initial-state memory, quantified by $\Delta r$, and
the effective temperature $\Teff$.  The distance to the thermal state
at long times is modified slightly by choosing a very large
$\delta_B$, but it should be noted that $\omega$ is itself dependent
on the energy width of the state when this becomes large on the scale
of changes in the density of states.

Conversely, we can make $\delta_B$ so small that there is just one
initial bath eigenstate in the initial product state and virtually
identical behaviour to Fig.~\ref{thermvars_W} is seen when $\lambda
\apprge 1$.  However, for smaller values of $\lambda$, fluctuations
appear as a function of $\lambda$, thus necessitating a finite
$\delta_B$.

\subsection{Eigenstate Thermalisation}
\label{sec:eigentherm}

We now discuss our results in relation to the eigenstate
thermalisation hypothesis (ETH). As discussed in
Section~\ref{sec:thermdef}, this hypothesis requires the eigenstate
expectation values of the subsystem projection operator,
$P_s$, (defined in Eq.~\eqref{eq:r_eev}) 
to depend only weakly on the exact choice of the
eigenstate $\ket{A}$. 

We expect ETH to be valid in the regime where we found thermalisation
in the previous sections --- for $U=J$, this regime covers a wide
range of coupling strengths, $0.1 \apprle \lambda \apprle 2$, with 
$\lambda\sim 1$ exhibiting behaviour closest to the canonical picture of 
thermalisation. So, we
will now study the dependence of eigenstate projections
$\bra{A}P_s\ket{A}$ on the coupling strength at $U=J$.
We will focus on the projection
on to the ground state of the subsystem in the ($\nparticle_s=2$, $s^z_s =0$)
sector at $U=J$.

\begin{figure}[hptb]
 \begin{centering}
   \includegraphics[scale=0.7]{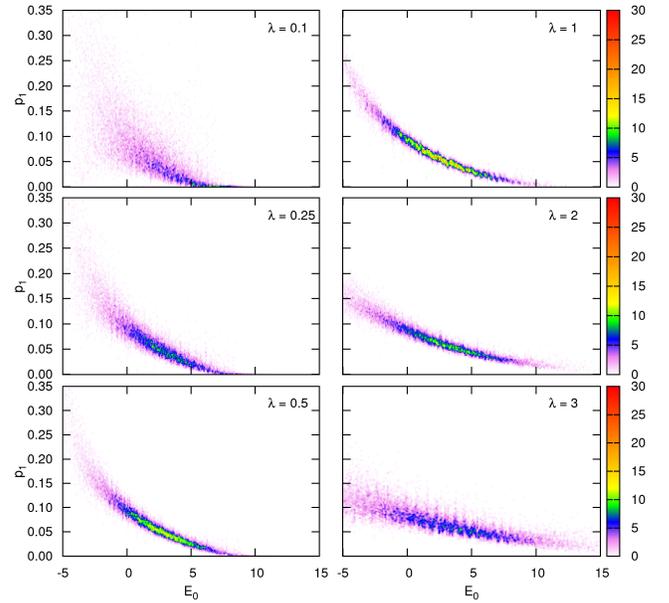}
   \caption{Histograms (plotted as colour scale) 
     of eigenstate projections $p_1=\bra{A}P_1\ket{A}$
     on to the two-site subsystem ground state, 
     for different subsystem-bath coupling $\lambda$.  
     ($U=J=1$, $\delta_B=0.5J$.)}
   \label{proj}
 \end{centering}
\end{figure}

`Perfect eigenstate thermalisation' corresponds to the projection
values forming a smooth quasi-continuous function of composite
eigenenergy $E_A$. When this occurs, complete independence of the
initial subsystem state exists. 
Fig.~\ref{proj} shows histograms of $\bra{A}P_{s=1}\ket{A}$.
We see that
there is some scatter in $\bra{A}P_1\ket{A}$ for different eigenstates
$\ket{A}$ that are close together in energy.  There is the least scatter
when the subsystem is closest to the canonical thermal state (small
$\Delta r$ and $\sigma_\omega$) at $\lambda \simeq 1$ for $U=J$.
Greater scatter in the values of $\bra{A}P_1\ket{A}$ is found when the
system starts to lose thermalisation (by our other measures of
thermalisation), at small $\lambda \apprle \lambdathermal=0.1$ and at
large $\lambda\apprge 2$.

\begin{figure}[hptb]
 \begin{centering}
   \includegraphics[scale=1.35]{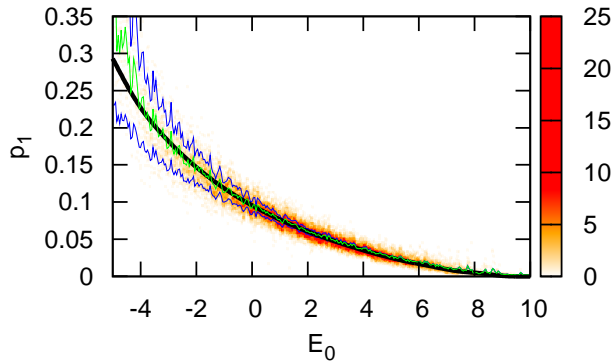}
   \caption[The same histogram of eigenstate projections
   $\bra{A}P_1\ket{A}$, from Fig.~\ref{proj} with $\lambda
   = 0.5$ is shown with a modified scale, which allows comparison with
   the thermal values $\omega_{11}$ and the mean eigenstate
   expectation-value from each vertical array of histogram bins to be
   shown clearly.]{The same histogram of eigenstate projections
     $p_1=\bra{A}P_1\ket{A}$, from Fig.~\ref{proj} with
     $\lambda = 0.5$ is shown with a modified colour scale, comparing
     the thermal values $\omega_{11}$ (black
     line) with the mean eigenstate projection from each
     vertical array of histogram bins (green line).  
     Positions of one standard
     deviation ($\spreadproj$) either side of the mean of the
     eigenstate projection are also shown (blue lines).}
   \label{projcomp}
 \end{centering}
\end{figure}

Let us examine the case of $\lambda=0.5$ at $U=J$ more closely. This shows
little scatter, and hence good eigenstate thermalisation, 
over a wide range of energies.
We quantify the extent to which eigenstate thermalisation holds by
measuring the mean and the standard deviation, $\spreadproj$, 
of the scattered values in each vertical column of
histogram bins on the plot in Fig.~\ref{proj}.  
This is computed using values within an energy window of
$0.5J$.
(This reduces the fluctuations in the measured $\spreadproj$.)
Fig.~\ref{projcomp} shows the positions
of the mean values of $\bra{A}P_1\ket{A}$
and the positions of $\pm\spreadproj$ 
from the mean. We see that there is indeed good agreement between the mean
eigenstate projections at any particular energy $E_0$ and the canonical
thermal value $\omega_{11}$ as defined in \eqref{eq:canonstate}.

\begin{figure}[htbp]
 \begin{centering}
   \includegraphics[scale=1.35]{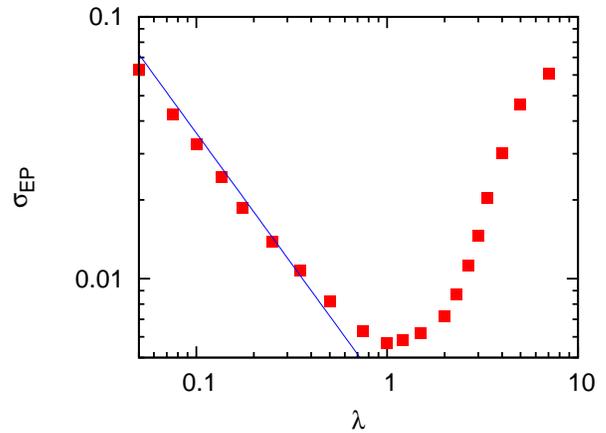}
   \caption[The spread of projection values, $\spreadproj$, as
   a function of subsystem-bath coupling strength $\lambda$.]{The
     spread of projection values, $\spreadproj$, as a function
     of subsystem-bath coupling strength $\lambda$ (hollow squares).
     The values of $\spreadproj$ were found by averaging over
     composite energies in a window of width $0.5J$ centered on
     $E_0=1.77$. Solid line: theoretical estimate
     \eqref{eq:spreadproj} 
     for small $\lambda$ 
     illustrating the scaling $\spreadproj \propto
     \lambda^{-1}$.}
   \label{projwidths}
 \end{centering}
\end{figure}

Let us now study the departure from eigenstate thermalisation,
measuring it by the increase in the scatter in the eigenstate
projection values. To reduce any bias due to changes in the density of
composite states with $\lambda$, we used eigenstates at the energy
$E_0=1.77J$ which is near the maximum in the density of states for all
coupling strengths considered here. The results are presented in
Fig.~\ref{projwidths}.  First of all, we observe that a minimum in
$\spreadproj$ indeed occurs over the same range of coupling strengths
where other measures of thermalisation also show that the subsystem
is close to a canonical thermal state.  Secondly, we find that, as the
subsystem departs from eigenstate thermalisation at low coupling
strengths, the increase in $\spreadproj$ with decreasing $\lambda$
obeys the relationship
\begin{equation}
\spreadproj \propto \frac{1}{\lambda} \quad\mbox{for $\lambda\apprle 1$.} 
\label{eq:sigmaEPscaling}
\end{equation}
We will discuss this scaling in Section \ref{sec:ETHscaling}. 

We also lose eigenstate thermalisation if we increase
the coupling strength to $\lambda\gg 1$. As discussed before, we believe that
this is a particular feature of our model where the properties of the
coupling dominate the Hamiltonian. 

Finally, recall that we found in Section~\ref{sec:couplingstrength}
that, as $\lambda$ is increased beyond unity at $U=J$, the steady
state of the subsystem departs from the canonical thermal state but
the RDM follows a good fit to the Boltzmann form. This seems to
indicate that the subsystem is in an effective thermal state that is
non-canonical. We can see an indication of this crossover regime in
Fig.~\ref{proj} for $1\le\lambda\le 3$, where the scatter in the
eigenstate projections is still relatively low, but the mean
eigenstate projections as a function of $E_0$ depart
significantly from the canonical thermal value $\omega_{11}$, in
contrast to the case at $\lambda=0.5$ (Fig.~\ref{projcomp}).

In summary, we have shown that eigenstate thermalisation
holds and agrees well with other measures of thermalisation. 
We will demonstrate in Section~\ref{sec:ETHscaling} that
the statistical behaviour of the eigenstate projections 
$\bra{A}P_s\ket{A}$ is consistent with a simple model of the eigenstates
$\ket{A}$ as random vectors in the basis of the subsystem-bath 
product states $\ket{sb}$.

\section{Eigenstate overlaps and the Eigenstate Thermalisation
  Hypothesis}
\label{sec:eth}

In Section \ref{sec:eigentherm}, we showed that the eigenstate
thermalisation hypothesis holds over a wide range of parameters for
our Hubbard-model system.  Deutsch~\cite{Deutsch1991} and
Srednicki~\cite{Srednicki1994} have suggested that
eigenstate thermalisation occurs if the composite system is quantum
chaotic. This was demonstrated theoretically for weak subsystem-bath
coupling using results for the eigenstates of generic (random)
Hamiltonians. In this section, we summarise these arguments and 
demonstrate numerically their agreement with our results for our
Hubbard-model lattice.
We will then use this framework to explain the
dependence of eigenstate thermalisaton on the coupling strength
discussed in section \ref{sec:eigentherm}, namely 
the scaling of the spread of eigenstate
projections, $\spreadproj$, with coupling strength $\lambda$ for $\lambda<1$.

\subsection{The Overlap Distribution and the Local Density of States}
\label{sec:ldos}

To be more specific, eigenstate thermalisation is concerned with the
eigenstate projections
$\bra{A}P_s\ket{A} = \sum_b \inner{A}{sb}\inner{sb}{A}$.
So, we need to understand the overlap $\inner{sb}{A}$ of the
eigenstates $\ket{A}$ of the composite system at non-zero coupling
with the eigenstates $\ket{sb}$ of the decoupled system at
$\lambda=0$.  As for the eigenstate projections $\bra{A}P_s\ket{A}$,
the overlaps will fluctuate if we change $E_A$ or $E_{sb}$. However,
we can study averages over energy windows that are narrow on the scale
of variation in the density of states but contain enough states to
smooth out fast fluctuations. 

The overlaps themselves are not invariant under a global gauge
transformation and so should have mean zero. Let us consider first the
squared overlap $|\inner{A}{sb}|^2$ whose average is the variance of
the overlaps. This can be interpreted as the
weight of the product state $\ket{sb}$ at energy $E_{sb}$ in the
decomposition of the eigenstate $\ket{A}$ at energy $E_A$ using all
the product states as the basis.  In this picture of the composite
eigenstate in energy space, $|\inner{A}{sb}|^2$ is called the `local
density of states'.  

We now discuss some known results for the local density of states.
For a coupling $\lambda V$ between subsystem and bath, we expect that
an eigenstate $\ket{A(\lambda)}$ at $E_A$ will consist mainly of
product states $\ket{sb}$ with energies $E_{sb}$ close to $E_A$.  If
the coupling is not strong ($\lambda < 1$), the energy range should
scale with the strength of the coupling matrix elements
$\lambda|\bra{s_As_B}V\ket{sb}|$ where $\ket{s_Ab_A}$ is the product
state corresponding to $\ket{A}$ in the limit $\lambda\to 0$.  To
leading order in $\lambda$, this can be written as $\lambda|\bra{A}
V\ket{sb}|$.  In this weak-coupling regime, we expect that the density
of state of the bath spectrum is nearly constant over this
range. Then, the mean value, $\overline{|\inner{A}{sb}|^2}$, should be
a strong function of the energy difference $\Delta E_{Asb} = E_A -
E_{sb}$, but has only a weak dependence on $E_A$ and $E_{sb}$
separately.  For a generic random coupling, we expect a Lorentzian
form in the dependence on the energy difference:
\begin{equation}
\label{eq:LDOS}
\begin{aligned}
\sigma^2_{Asb} &\equiv \overline{|\inner{A(\lambda)}{sb}|^2} 
=  \frac{\lambda^2 \overline{|\bra{A}V\ket{sb}|^2}}{W_L^2 + (E_A-E_{sb})^2} \\
W_L &= \pi \lambda^2 g(E_{sb})  \overline{|\bra{A}V\ket{sb}|^2}\,.
\end{aligned}
\end{equation}
where $g$ is the density of states of the composite system evaluated
at the total energy $E_{sb}$, taken to be approximately constant over
the energy width of this Lorentzian so that $g(E_A)\simeq g(E_{sb})$
in this range of energy. It may be related to
straightforward perturbation theoretic
results~\cite{Kottos2001,*Cohen2000,*Hiller2006} to second order in
$\lambda$.  This result was originally
established~\cite{Wigner1955,*Wigner1957} over half a century ago for a
specific model of random coupling.  This result was later shown to
hold more generally. However, we note that this result does not take
into account the specific case of coupling a bipartite
system. Moreover, Eq.~\eqref{eq:LDOS} is only strictly accurate in the
general case for energy differences where $|E_A - E_B| > W_L$.  At
smaller energy scales, non-perturbative mixing occurs and it is no
longer possible to associate eigenstates with a specific unperturbed
state.  However, the presence of non-perturbative mixing between
states separated by less than $W_L$ leads us to make the assumption
that structures in the coupling matrix, such as elements which are
identically zero because of the precise nature of the Hubbard-model
coupling, are washed out by this mixing.  We should also
stress that this Lorentzian form is not expected to hold when $\lambda
\sim 1$. However, this is sufficient for us to use this form in the
discussion of this section where we are concerned with the behaviour
of the system at weak coupling. We note that, at stronger coupling, we
find a Gaussian form for the local density of states~\footnote{In preparation.}.

\begin{figure}[!thb]
 \begin{centering}
   \includegraphics[scale=1.12]{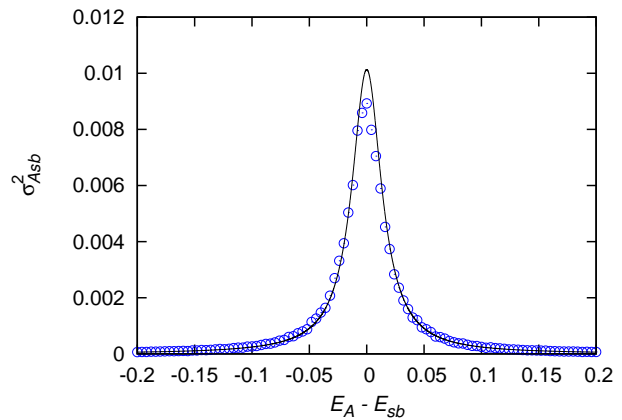}
   \caption{Local density of states, $\sigma^2_{Asb}$, as a function
   of $\Delta E  = E_A - E_{sb}$ for the Hubbard model with $U=J=1$,
     at weak coupling $\lambda=0.1$. The averaging uses the overlaps of
     all eigenstates $\ket{A}$, with 
     $\ket{s}_S = \ket{\!\uparrow,\uparrow}$ and
     $\ket{b}_B$ selected within energy $J$ from the centre of the
     bath spectrum. Solid line: Lorentzian with width $W_L$ as given
     by \eqref{eq:lorentzianwidth}.}
   \label{LDOS}
 \end{centering}
\end{figure}

We can make an estimate for the width $W_L$ in the Lorentzian form
\eqref{eq:LDOS}. As discussed above, the overlap $\bra{A} V\ket{sb}$
can be approximated by $\bra{s_Ab_A} V\ket{sb}$ to leading order in
$\lambda$.  So, we see that its mean square value should be the mean
square value $\overline{V^2}$ of an element of the coupling matrix
$\bra{s'b'}V\ket{sb}$. We
will see in Eq.~\eqref{eq:matrixelement} in
Section~\ref{sec:sizescaling} that $\overline{V^2} \simeq J\Delta /2$
for an interacting system near half filling.  If we further
approximate the density of states $g(E_{sb})$ with the average level
spacing $\Delta$, we see that
\begin{equation}\label{eq:lorentzianwidth}
W_L \simeq \frac{\pi\lambda^2 J}{2}\qquad (\mbox{for\ }U\sim J)\,.
\end{equation}
The local density of states, $\sigma^2_{Asb}$, for the Hubbard model at $\lambda=0.1$ and
$U=J$ is shown in Fig.~\ref{LDOS}. It fits well to the Lorentzian form
\eqref{eq:LDOS} with $W_L$ given by \eqref{eq:lorentzianwidth}.

\begin{figure}[!thb]
 \begin{centering}
   \includegraphics[scale=0.675]{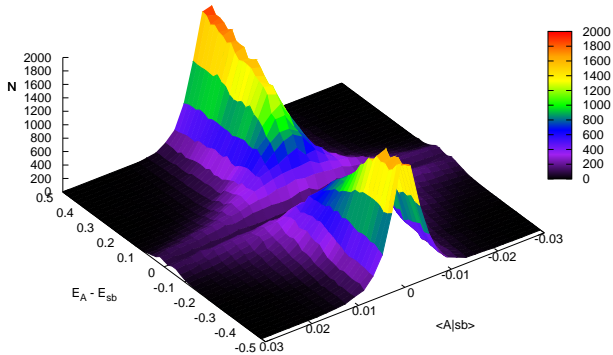}
   \caption{Two-dimensional histogram showing the distribution
     of the overlaps $\inner{A}{sb}$ as a function of the energy
     difference $E_A - E_{sb}$ for the Hubbard model with $U=J=1$,
     at weak coupling $\lambda=0.1$. The histogram includes the overlaps of
     all eigenstates $\ket{A}$, with 
     $\ket{s}_S = \ket{\uparrow,\uparrow}$ and
     $\ket{b}_B$ selected within energy $J$ from 
     the centre of the bath spectrum.  
     The histogram bin widths are 0.02 and 0.002 on the
     energy and overlap axes respectively.}
   \label{overlapdist_hist}
 \end{centering}
\end{figure}

We can also discuss the full distribution of the overlaps
$\inner{sb}{A}$. For a system with time reversal symmetry,
$\inner{sb}{A}$ can be constructed to be real. Our numerical results
for a system at $U=J$ and $\lambda=0.1$ are shown in
Fig.~\ref{overlapdist_hist}. This is a histogram using the overlaps
of all the eigenstates $\ket{A}$ with a subset of product states
$\ket{sb}$ where the bath states are within an energy $J$ of the
centre of the bath spectrum.  We can see that the
width of the distribution is a strong function of the energy
difference $\Delta
E_{Asb}$ between $\ket{A}$ and $\ket{sb}$. 
The widest distribution is found at $E_A=E_{sb}$.  In this
case, the states $\ket{sb}$ effectively form a random basis for the
eigenstates $\ket{A}$.

\begin{figure}[!hbt]
 \begin{centering}
   \includegraphics[scale=0.97]{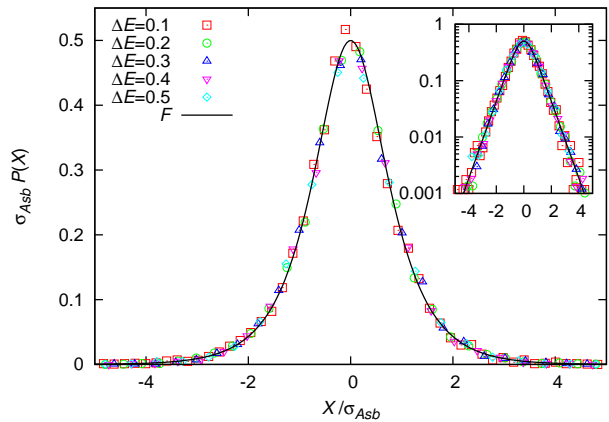}
   \caption{Distribution of overlaps $X=\inner{A}{sb}$ at weak
     coupling $\lambda=0.1$, scaled by the width $\sigma_{Asb}(\Delta
     E = E_A-E_{sb})$ at different values of $\Delta E$. ($J=1$; same
     system parameters as in Fig.~\ref{overlapdist_hist}.)  Solid
     line: hyperbolic secant distribution with zero mean and unit
     variance. Inset: Log plot of the same data.}
   \label{overlapdist_sech}
 \end{centering}
\end{figure}

The distributions appear to be controlled by a single variable, the
local density of states. In other words,
\begin{equation}
\label{eq:overlapdist_collapse}
P(X = \inner{sb}{A}) = \sigma^{-1}_{Asb} F(X/\sigma_{Asb})
\end{equation}
for a normalised distribution $F(u)$ with unit variance. 
This is demonstrated in Fig.~\ref{overlapdist_sech} for our data
at five different $\Delta E_{Asb}$.  The data have been scaled using the
expected width $\sigma_{Asb}$ given by \eqref{eq:LDOS} and
\eqref{eq:lorentzianwidth}. So, this data collapse contains no
adjustable parameters. The distribution $F(u)$ has an
excess kurtosis $\gamma = \langle u^4\rangle - 3\langle u^2\rangle^2
\simeq 2$ numerically. ($\gamma$ would be zero for a Normal
distribution.) In Fig.~\ref{overlapdist_sech}, we see that our data
are well approximated by a hyperbolic secant distribution 
which has an excess kurtosis of 2:
\begin{equation}
\label{eq:distsech}
F(u)  = \frac{1}{2\cosh(\pi u/2)}\,.
\end{equation}
We should point out that the data collapse to this distribution fails at
strong coupling. This may be due to the fact that the width of the
distribution $\sigma_{Asb}$ becomes large enough that each eigenstate
$\ket{A}$ involves bath states in a wide range of energies over which the bath
density of states varies significantly.

We note that our form for the overlap distribution differs from what may be
expected from random matrix
theory~\cite{Berry1977,Deutsch1991,Srednicki1994} for similar types of
overlaps which suggests that they should follow a Normal distribution
at weak coupling. This indicates that the overlap distribution may depend
details of the coupling Hamiltonian or details of random matrix
ensemble.

Accepting the distribution \eqref{eq:overlapdist_collapse} as the
distribution for the overlaps, the distribution $P^\pldos$ for the
squared overlaps $|\inner{sb}{A}|^2$ can be derived:
\begin{equation}\label{eq:PLDOS}
    P^\pldos \left( |\inner{A}{sb}|^2 = Y \right)
    = \frac{F(\sqrt{Y}/\sigma_{Asb})}{\sigma_{Asb} \sqrt{Y}} \,.
\end{equation}
which has a mean of $\sigma^2_{Asb}$ and a variance of
$(2+\gamma)\sigma^4_{Asb}$. 
In the case where the overlap distribution is so wide
that $\ket{A}$ is effectively a random vector in the basis of
$\ket{sb}$, we have a Porter-Thomas distribution for the local density
of states.

To summarise, we have shown that our numerics agree with results for
the local density of states arising from generic random
Hamiltonians. This controls the overlap distribution. 
We point out that in this simple
picture of the statistics of the overlaps, any correlations between
different eigenstate overlaps are implicitly neglected.  We will now
proceed to understand eigenstate thermalisation in terms of this
simple picture of eigenstate overlaps~$\inner{ab}{A}$.

\subsection{Scaling of Eigenstate Thermalisation with Coupling Strength}
\label{sec:ETHscaling}

In Section~\ref{sec:eigentherm}, we found that the degree of
eigenstate thermalisation improves upon increasing the strength of the
subsystem-bath coupling. In the previous section, we have seen
\eqref{eq:LDOS} that, in parallel, increasing the coupling strength
broadens the local density of states.  We expect that, as the
distribution of overlaps $\inner{sb}{A}$ broadens such that more basis
states $\ket{sb}$ participate in each eigenstate, the fluctuations in
the projection values $\bra{A}P_s\ket{A}$ will be reduced in
accordance with the law of large numbers.
We will show this to be the case
and find the $\lambda$-dependence for the spread of eigenstate
projections $\spreadproj$, found numerically in
Section~\ref{sec:numerics}.

From its definition \eqref{eq:r_eev}, the projection operator $P_s$ sums
over all bath states. In our model of the overlaps, $\bra{A}P_s\ket{A} = \sum_b
|\inner{sb}{A}|^2$ is a sum over many independent variables. 
Its mean and variance are given by
\begin{equation}\label{eq:EPstats}
\mu_\EP = \sum_b \sigma^2_{Asb}\,,\quad
\spreadproj^2 = (2+\gamma)\sum_b \sigma^4_{Asb}\,.
\end{equation}
For the full distribution of these quantities, see
Appendix~\ref{sec:appendix} which applies the central limit theorem to
our model distribution for the overlaps. 

First, we consider the mean $\mu_{\EP, As}$ for a given subsystem state
$s$. From our model $\eqref{eq:LDOS}$, this is the sum over all bath
states (with the necessary spin and particle number) using a Lorentzian window of
energy centered at $\egybath_b = E_A-\egysys_s$. In other words,
the answer should be proportional to the bath density of states at
$E_A-\egysys_s$. Furthermore, we have the normalisation condition
$\inner{A}{A}=\sum_{sb}\sigma^2_{Asb} = \sum_s \mu_{\EP,As}=1$. So, at fixed $A$,
$\mu_\EP$ should give the normalised probability of finding the subsystem in
state $s$ according to the Gibbs distribution:
\begin{multline}\label{eq:rmt_therm}
\mu_\EP =\sum_b \sigma_{Asb}^2 
= \frac{\sum_b \sigma_{Asb}^2}{\sum_s \sum_b \sigma_{Asb}^2}\\ 
\simeq 
\frac{g_B(E_A-\egysys_s,\Nparticle - \nparticle_s, S^z -
  s^z_s)}{g(E_A,\Nparticle,S^z)}
= \omega_{ss}(E_A) 
\end{multline}
where $E_A$, $\Nparticle$ and $S^z$ are the energy, number and spin of the
state $\ket{A}$,
$g_B(\egybath_b,\nparticle_b,s^z_b)$ is the density of bath states with energy
in an interval about $\egybath_b$, $\nparticle_b$ particles and spin $s^z_b$.
Thus, we see that a simple model of the eigenstate overlaps gives the
canonical thermal distribution~\cite{Deutsch1991,Srednicki1994,Tasaki1998}. 

Next we address the spread of the projection values $\spreadproj$.
Using the Lorentzian form \eqref{eq:LDOS} for the local density of
states with width $W_L$, and following the same approximations as
above, the sum over $\sigma_{A,sb}^4$ is:
\begin{align}
\sum_b&\sigma^4_{A,sb} =\int 
\frac{g_B(\egybath_b)  W^2_L/\pi^2 g^2(E_A)}{[W_L^2 + (E_A-\egysys_s -
    \egybath_b)^2]^2}d\egybath_b \notag\\
&\simeq \frac{g_B(E_A-\egysys_s)}{g^2(E_A)} \int\!\!
\frac{(W^2_L/\pi^2) d\egybath_b }{[W_L^2 + (E_A-\egysys_s -
    \egybath_b)^2]^2}\notag\\
&=  \frac{\omega_{ss}(E_A)}{2\pi W_L g(E_A)}\,.
\label{eq:sigma4sum}
\end{align}
This means that the spread of the projection values is given by:
\begin{equation}
\spreadproj =\sqrt{\frac{(2+\gamma) \omega_{ss}(E_A)}{2\pi W_L g(E_A)}} 
= \frac{1}{\pi\lambda}\sqrt{\frac{(2+\gamma) \mu_{\EP,As}}{Jg(E_A)}}
\label{eq:spreadproj}
\end{equation}
where we have used our estimate \eqref{eq:lorentzianwidth} for the
Lorentzian width $W_L$. Note that $1/g(E_A)$ is of the order of the average
level spacing $\Delta$. 

Therefore, we find that $\spreadproj$ is proportional to $\lambda^{-1}$
as shown numerically in Fig.~\ref{projwidths}.  Moreover, we see that
$\spreadproj \sim \sqrt{\Delta}$ so that the
fluctuations in the projection values are
small for large systems, in accordance with the law of large numbers
for a quantity that is a sum over many states.

We stress that the above results hold for any distribution of
matrix elements $\inner{A}{sb}$, of sensible form, where the central
limit theorem applies.  Furthermore, the result~\eqref{eq:rmt_therm}
holds quite generally for any sensible form of $\sigma^2_{Asb}$ which
is a function of $E_A-E_{sb}$ with a peak at $E_A-E_{sb}=0$.
Although we have not derived it here explicitly, it should also be
noted that the off-diagonal elements of the reduced density matrix,
which were found to be virtually zero numerically, are expected to be
zero from this model of eigenstate overlaps.  Indeed the mean values
of the eigenstate expectation-values for off-diagonal elements are
clearly zero due to the random sign of eigenstate overlaps.

\section{Crossover to Thermalisation}
\label{sec:sizescaling}

In this section, we will try to understand the onset of thermalisation
using simple theoretical arguments.  In particular, we explore the
effects of system size on the thermalisation for the cases of
interacting and non-interacting fermions in the Hubbard model.  The
extent to which the effects of system size on thermalisation may be
seen numerically is limited, as discussed in Section~\ref{sec:thsc}.
We proceed to identify a minimum coupling strength $\lambdanp$ below
which thermalisation cannot occur. 

To begin to see thermalisation
requires the coupling Hamiltonian $\lambda V$ to be big enough to mix
the $\lambda=0$ eigenstates non-perturbatively. We will then compare
this theoretical estimate with our numerical results.

For small coupling strength ($\lambda \ll 1$), the overlap
between an eigenstate $\ket{A}$ and a subsystem-bath product state $\ket{sb}$
takes the form
\begin{multline}
\inner{sb}{A} \simeq  \delta_{s_As}\delta_{b_Ab} 
+ \lambda\left[\frac{\bra{sb} V\ket{s_Ab_A}}{E_{s_Ab_A}-E_{sb}}\right]\\
+\!\! \lambda^2\bigg[
\sum_{s'b'\neq s_Ab_A}\! 
\frac{\bra{sb}V\ket{s'b'}\bra{s'b'}V \ket{s_Ab_A}}{(E_{s_Ab_A}-E_{sb})(E_{s_Ab_A}-E_{s'b'})} \\
- \frac{1}{2} \frac{|\bra{sb} V \ket{s_Ab_A}|^2}{(E_{s_As_B} - E_{sb})^2}\bigg]
\end{multline}
to second order in $\lambda$, where the state
$\ket{s_Ab_A}$ is the composite eigenstate $\ket{A}$ to zeroth order
in $\lambda$. (Note that $V$ has no diagonal elements in this basis.)  
The threshold for non-perturbative mixing may be
considered to be met when the second-order term equals the first order
term in magnitude. Note that the bath states $\ket{b}_B$ coupled by $V$ have
different quantum numbers from the given state $\ket{b_A}_B$ so that
there is no level repulsion between $\ket{b}_B$ and $\ket{b_A}_B$.  We
expect the nearest bath state is on average $\Delta_B/4$ away in
energy.  Generically, this occurs around a coupling strength
$\lambdanp$ which we define by
\begin{equation}
\lambdanp\left[\overline{V^2}\right]^{\frac{1}{2}} =
\frac{\Delta_B}{4}\,,
\label{eq:lambdanp_def}
\end{equation} 
where $\Delta_B$ is the bath level spacing and $\overline{V^2} =
\overline{|\bra{s'b'}V\ket{sb}|^2}$ is the typical magnitude of the
square of a coupling matrix element. We will estimate these below for
interacting and non-interacting systems.

The quantity $\lambda_{\text{np}}$ should be the coupling strength at
which one starts to see a departure from complete memory of the
initial state at long times.  We therefore expect that this quantity
should be similar to the quantity $\lambdathermal$, introduced in
Section~\ref{sec:numerics}, which measures the crossover from the
non-thermalised regime to thermalisation.  Note that $\lambdathermal$
has been defined with an arbitrary choice of a threshold for
$\sigma_\omega$ at 25\%. Its actual value will change with the
specific criterion chosen to mark this threshold.  However, one can
use the data from Fig.~\ref{thermsites} to show that the relative
values for $\lambdathermal$ for different system parameters are
approximately the same for a range of choice of thresholds.  So, it is
reasonable to discuss a relationship between $\lambdanp$ and
$\lambdathermal$. In particular, it is expected that the two
quantities should be proportional to each other for a given subsystem
size.


The rest of this section is dedicated to
understanding the scaling of $\lambda_{\text{np}}$ with system size.
First we will consider the case of finite interactions $U\sim J$
before, in the subsection following, discussing the case of virtually
free fermions where $U \ll J$.

\subsection{System-Size Scaling for Interacting Fermions}
\label{sec:SystemSizeScalingInteracting}

\begin{figure}[htbp]
 \begin{centering}
   \includegraphics[scale=1.5]{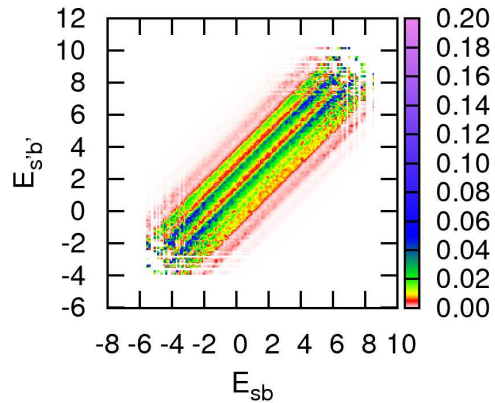}
   \caption[A plot of the magnitude of coupling matrix elements
   linking the $\nparticle_s=3,\,s^z_s=\frac{1}{2}$ and $\nparticle_s=2,\,s^z_s=1$
   subsectors for the Hubbard model with $U=J=1$.]{
     Coupling matrix elements linking the $\nparticle_s=3,\,s^z_s=\frac{1}{2}$
     and $\nparticle_s=2,\,s^z_s=1$ subsectors for the Hubbard model with
     $U=J=1$. Colour scale indicates the magnitude of the matrix
     elements. 
     The banded diagonal structure is typical for the
     coupling of all subsystem states.  Sizeable
     matrix elements lie within a band of width $4J=4$, with some very
     small matrix elements lying outside of the band due to the finite
     interaction strength $U$.}
   \label{VUeqJ}
 \end{centering}
\end{figure}

We will now deduce the scaling of $\lambdanp$ with system size for the
Hubbard model with interactions $U\sim J$.  To find the theoretical
scaling of $\lambdanp$ with system size requires a knowledge of the
scaling of both the energy spacing between coupled states and the
scaling of the magnitude of the typical matrix elements
$\bra{sb}V\ket{s'b'}$ with system size.  A characteristic submatrix of
the coupling matrix $\bra{sb}V\ket{s'b'}$, with $s$ and $s'$ fixed, is
shown in Fig.~\ref{VUeqJ} for the Hubbard model with interaction
strength $U=J$.  The non-zero elements of the coupling matrix form a
band. This can be explained by the single-particle nature of the
coupling.  In the limit of zero interactions, the coupling involves
a single particle hopping into, or from, 
one of the single-particle states in the bath.
Therefore, the full width, $2W$, for bath states into which a particle may hop
is $4J$, the single-particle bandwidth.  The presence of interactions
preserves the banded structure of the coupling matrix and, provided $U
\apprle J$, the banded matrix is not significantly broadened beyond
$4J$.  However, when $U \sim J$, the details of single-particle bath
states are blurred as the single-particle quasiparticle weight
is significantly reduced from unity.

First of all, we estimate the magnitude of a matrix element
$\bra{sb}V\ket{s'b'}$. To keep the description
straightforward, we consider the case of exactly half filling. We will
compute this from the average for the sum of all the squared matrix
elements ${\rm Tr} V^2 = \sum_{ss'bb'}|\bra{sb}V\ket{s'b'}|^2$.
The calculation of this trace can be found in Appendix \ref{sec:tracevsq}.
We find:
\begin{equation}
\label{eq:trVsq}
\Tr(V^2) = 2 \nstate J^2\,,
\end{equation} 
where $\nstate$ is the dimension of the Hilbert space of the composite
system. We now need to count the number of non-zero matrix elements in the
coupling matrix $\bra{sb}V\ket{s'b'}$. Since the coupling involves the
hopping of a single particle of a given spin state between the
subsystem and bath, any given subsystem state $s$, 
will only have non-zero matrix elements with at most four other subsystem
states $s'$, corresponding to changing the particle number or spin by $\pm 1$.
So, there should be approximately $4\nstate_S$ such non-zero blocks in
the coupling matrix where $\nstate_S$ is the dimension of the subsystem
Hilbert space. Each block has a banded structure similar to the one shown
in Fig.~\ref{VUeqJ}. Note that the bath states, $b$ and $b'$, 
connected by $\bra{sb}V\ket{s'b'}$ belong to subsectors of the bath
spectrum with different quantum numbers. 
For a band of full width $2W$, the banded block
should have $\nstate_B(2W/\Delta_B)$ non-zero elements where $\Delta_B$ is
the average bath level spacing and $\nstate_B$ is the total number of bath
states in the bath subsector of a given number and spin. So, the total
number of non-zero elements in the coupling matrix $V$ is
approximately $4\nstate_S \nstate_B(2W/\Delta_B)\simeq 8\nstate_BW/\Delta$. 
Therefore, the mean squared value
of each coupling matrix element, $\overline{V^2}$, is
\begin{equation}
\label{eq:matrixelement}
\overline{V^2} 
\simeq \frac{2 \nstate J^2}{(8W\nstate_S\nstate_B/\Delta_B)}
\simeq\frac{J^2 \Delta_B}{4W}\simeq \frac{J\Delta_B}{8}\,,
\end{equation}
using $2W = 4J$ which is valid for the case in Fig.~\ref{VUeqJ} where
$U\sim J$ as discussed above. So, from \eqref{eq:lambdanp_def}, we see
that non-perturbative mixing occurs when $\lambda$ reaches
\begin{equation}
\lambdanp = \sqrt{\frac{\Delta_B}{2J}}\simeq 
\sqrt{\frac{\nstate_S\Delta}{2J}}\,.
\label{eq:lambdanp_Delta}
\end{equation}
For the purposes of understanding how this threshold scales with
system size, we have approximated the bath level spacing as a
simple multiple of the average level spacing $\Delta$ of the composite
system: $\Delta_B\approx \nstate_S\Delta$.

We have arrived at this condition using simple
arguments based on perturbation theory. A
similar criterion can be obtained using our results for eigenstate
thermalisation. In terms of the eigenstate projection values,
the system does not thermalise if the spread of the projection values,
$\sigma_\EP$, becomes comparable to the mean $\mu_\EP$. We have shown
in \eqref{eq:rmt_therm} that the latter gives the canonical state
$\omega$ which is of the order of $1/\nstate_S$ where $\nstate_S$ is 
the number of states in the subsystem.
Thus, from \eqref{eq:spreadproj}, the condition that
$\spreadproj<\mu_\EP$ for a given total energy $E_0$ and a subsystem
state $s$ becomes the criterion that 
$\lambda > \lambdaeth(E_0,\egysys_s)$ where
\begin{equation}
\lambdaeth = 
\sqrt{\frac{2+\gamma}{Jg_B(E_0-\egysys_s)}}
\label{eq:lambda_eth}
\end{equation}
Since $1/\Delta_B$ is simply the average of $g_B$ over the bath sector, 
we see that the thresholds $\lambdanp$ and
$\lambdaeth$, based on different criteria, 
describe essentially the same crossover.
$\lambdaeth$ is larger than $\lambdanp$ as might be expected since the
latter marks the loss of memory of the initial state while the latter
marks the onset of the canonical thermal state. 

It remains to establish how the level spacings
$\Delta$ depends on the system size.  Assuming
the cosine dispersion for a tight-binding band and neglecting the
broadening due to finite $U$, the many-body bandwidth for $L$ sites is
$8JL/\pi$ at half filling.  This is found approximately by
finding the maximum and minimum composite energy eigenvalues, $\pm
4JL/\pi$, by summing the energies of the $L/2$
highest and $L/2$ lowest single-particle eigenstates.
Therefore, the mean level spacing is
\begin{equation}
\Delta = \frac{8JL}{\nstate\pi}\,.
\label{eq:del}
\end{equation}
At half filling, the Hilbert-space dimension is
\begin{equation}
  \nstate = \left[\frac{L!}{\left(\frac{L}{2}\right)!
      \left(\frac{L}{2}\right)!}\right]^2 \,.
\label{eq:vsq}
\end{equation}
Therefore, using \eqref{eq:lambdanp_Delta} and Stirling's
approximation for large $L$, we find that
the threshold for the loss of memory of the initial state which allows
the onset of thermalisation occurs at
\begin{equation}
\lambdanp(U\sim J) =  4\sqrt{2} L\,2^{-L}\,.
\label{eq:lambdanp_size}
\end{equation}
Reassuringly, $\lambdanp$ tends to zero as $L$ tends to infinity so
that, for baths in the thermodynamic limit, arbitrarily small
couplings lead to thermalisation, as we expect~\cite{Feynman}.  For a
lattice of nine sites we estimate this sets $\lambdanp \simeq 0.054$.

\begin{figure}[!hbt]
\begin{centering}
\includegraphics[scale=1.3]{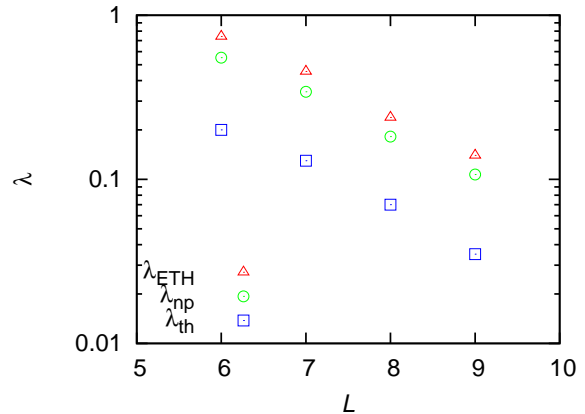}
\caption[A comparison of $\lambdathermal$, $\lambdanp$ and $\lambdaeth$ for
different system sizes $L$ when $U=J$.]{A comparison of
  $\lambdathermal$, $\lambdanp$ and $\lambdaeth$ 
  for different system sizes $L$ when
  $U=J$.  $\lambdathermal$ is set as the coupling strength for
  which $\sigma_{\omega}$ falls to
  0.25 for composite energies $E_0$ in the centre of the band.
  $\lambdanp$ is found using \eqref{eq:lambdanp_Delta} with
  \eqref{eq:del}.
  $\lambdaeth$ is obtained  from \eqref{eq:lambda_eth}. }
\label{npth}
\end{centering}
\end{figure}

We now compare $\lambdanp$, for different lattice sizes, with
$\lambdathermal$.  To allow easy comparison between different system
sizes, the composite energy in the centre of the band will be
considered in each case.  As already discussed, the value of
$\sigma_{\omega}$ where the value $\lambdathermal$ is recorded is
somewhat arbitrary.  However, we find good agreement between
$\lambdathermal$ and $\lambdanp$, as is shown in Fig.~\ref{npth}, with
$\lambdathermal$ defined using a threshold of $\sigma_\omega=25\%$.
If thresholds other than $\sigma_{\omega}$=25\% are considered, we
find that $\lambdathermal$ changes approximately by a multiplicative
constant for all $L$.  Therefore, the good agreement between the
explicit values for $\lambdathermal$ and $\lambdanp$ is not a
remarkable feature of Fig.~\ref{npth}.  However, that the two
quantities are found to scale in virtually the same way for the
limited numerical data available provides numerical evidence to
support the system-size scaling of
thermalisation~\eqref{eq:lambdanp_size} derived above.

\subsection{System-Size Dependence for Non-Interacting Fermions}
\label{sec:UllJ}

\begin{figure}[!htbp]
 \begin{centering}
   \includegraphics[scale=1.5]{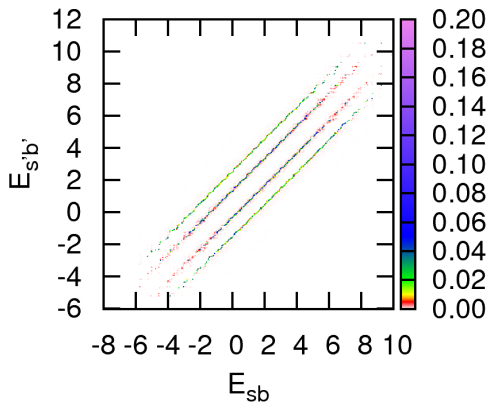}
   \caption[A plot of the magnitude of coupling matrix elements
   between two different subsystem states.]{Coupling matrix elements linking the $\nparticle_s=3,\,s^z_s=\frac{1}{2}$
     and $\nparticle_s=2,\,s^z_s=1$ subsectors for the Hubbard model with
     ${U=0.01J=0.01}$.  Colour scale indicates the magnitude of the matrix
     elements.  The banded diagonal structure is typical for
     the coupling of all subsystem states.  The
     finite matrix elements lie within a band of width $4J$ but, in
     contrast to Fig.~\ref{VUeqJ}, the matrix appears striped.}
   \label{VU0}
 \end{centering}
\end{figure}

For the case of almost free fermions ($U\ll J$), thermalisation was
not seen in the nine-site Hubbard ring.  We now repeat the argument
above for the case of negligible $U$.  The major difference from the
case of $U\sim J$ above
is the structure of the coupling matrix.  Shown in Fig.~\ref{VU0} is
the striped form of the coupling matrix.  Without interactions, the
bath eigenstates are simply Slater determinants of free-fermion
single-particle states.  Therefore, for each spin, the coupling
Hamiltonian has non-zero matrix elements only at energies
corresponding to the $L-2$ single-particle bath states for each spin.

When considering the threshold for non-perturbative mixing, it must
now be noted that coupled states differ not by the level spacing
$\Delta$, but by the bath single-particle level spacing $\Delta_1$,
where $\Delta_1 \approx 4J/(L-2)$. The
magnitude of $\Tr{V^2}$, as given by \eqref{eq:trVsq}, is independent
of $U$. Therefore, using \eqref{eq:lambdanp_Delta} with $\Delta$
replaced by $\Delta_1$, 
we see that $\lambdanp$ for $U\ll J$ should be given by
\begin{equation}
\lambdanp(U\ll J) = 2\sqrt{\frac{2}{L-2}}\,.
\label{eq:lambdanp2}
\end{equation}
This yields a value $\lambdanp \simeq 1.1$ for the nine-site lattice.
It is therefore clear why initial-state independence is not seen for the
nine-site lattice when $U\ll J$:  the threshold for non-perturbative
coupling occurs at a coupling strength some 20 times bigger than for
the Hubbard model with interactions $U\sim J$. 
Eq.~\eqref{eq:lambdanp2} indicates that the $U\ll J$ case needs 
a lattice with 2700
sites in order that $\lambdanp$ falls to the same
value as for the $U\sim J$ interacting case.

We have argued that thermalisation for small systems occurs at smaller
system sizes for the interacting system compared to the
non-interacting system. We will now ask how the crossover from the
$U\ll J$ regime to the $U\sim J$ regime occurs. This should occur when
the width of the stripes in the coupling matrix elements seen in
Fig.~\ref{VU0} becomes comparable to the single-particle level spacing
$\Delta_1$. The stripe width should be the quasiparticle
decay rate due to interparticle collisions. 
At small $U$, the decay rate can be estimated using Fermi's golden
rule. The matrix elements are proportional to $U$ and the density of
single-particle states is proportional to $1/J$.
So, for single-particle energies far from the Fermi level so that
we can neglect effects from Pauli exclusion, the decay rate should
be $\sim U^2/J$. This becomes comparable to $\Delta_1$,
when $U$ reaches $\Uthermal \sim \sqrt{J\Delta_1} \propto 1/\sqrt{L-2}$.
This estimate give the scale for the interaction strength beyond which
we see thermalisation in the numerical results 
shown in Fig.~\ref{thermvars_U_sizes}. We have only four system sizes
and there is a strong even-odd
effect in the system size, making it difficult to verify our
prediction quantitatively.

\section{Experimental Implications}
\label{sec:expt}

Models such as the Hubbard model studied in this work can be
simulated readily using cold atoms in optical
lattices~\cite{Bloch2005}. Thanks to recent rapid progress in addressing single sites in optical
lattices \cite{Bakr2009,Sherson2010,Weitenberg2011,Endres2011}, models
similar to those we studied here can now in principle be implemented
and measured in systems of ultracold atoms trapped in optical
lattices. In particular, single-site imaging capability means that
atom occupation (albeit up to number modulo two) and the spin species
can be determined accurately at a few lattice sites that will form the
subsystem. This means the state of the subsystem can be probed
directly.  Single-site addressability means the subsystem and the bath
can be initialised with pure quantum states with well-defined number
and spin. Furthermore, instead of focusing the probe laser beam on a
single site, the laser can be aimed accurately (to within a tenth of
the lattice spacing \cite{Weitenberg2011}) between two neighbouring
sites, to tune the lattice potential locally and thus adjust the
coupling $\lambda J$ between the subsystem and bath. Both $\lambda \geq
1$ or $\lambda <1$ regimes can be accessed with a blue- or red-detuned
laser focussed between the sites.

We expect our findings to be seen for bath
state with a relatively well-defined energy (which overlaps bath
eigenstates only within a range of energies much smaller than the
many-body bandwidth), far from a strongly-correlated ground state.
We should point out that, although we focussed on a
specific Hamiltonian with specific initial conditions, we believe that
our results are generally applicable. For instance, we obtain similar
results for Bose and fermion Hubbard models. Also, although we have used an initial
bath state \eqref{eq:windowstate} consisting of bath eigenstates within a narrow
energy window, our results (Fig.~\ref{thermvars_W}) are not sensitive to
the width of this energy window. Our results should hold for initial states
spanning a larger window of bath energies which would be easier to
prepare experimentally. Our results also do not
change if we introduce random coefficients in the linear superposition
of bath eigenstates or change the shape of the window, consistent with the results of Ref~\cite{Linden2009}.

Moreover, experimental systems will be larger and contain many more
atoms than in our simulation. We have shown that the threshold $\lambdathermal$ in the
coupling strength is exponentially small in the system size for large
systems.  So, we believe
it would be possible to see thermalisation at smaller $\lambda$ in
experimental systems, even if the initial bath state is simply a single eigenstate.

We should also ensure that the time needed for thermalisation to be seen should be within the
lifetime of an optical lattice experiment, typically hundreds of
milliseconds or more. Our previous work~\cite{Genway2010} for systems
with $U\simeq J$ shows that the relaxation towards equilibrium occurs
with a relaxation rate $\sim \lambda^2 J$ for weak coupling (showing
exponential decay) and $\sim \lambda J$ when $\lambda \sim 1$ (showing
Gaussian decay).  (See Fig.~4 of~\cite{Genway2010}.)  For current
optical lattice experiments with $^{40}$K, using an optical lattice
laser wavelength of $1064$\,nm, the hopping matrix element $J$ is
approximately 380\,Hz for a laser strength $V_0$ of 5 times the recoil
energy ($E_R$), and $\approx 100$\,Hz for $V_0 = 10 E_R$.  Hence,
expected relaxation time scales when $V_0=5E_R$ range from about
$3$\,ms at $\lambda=1$ in the Gaussian regime, to about $30$\,ms at
$\lambda = 0.1$ in the exponential regime. (The corresponding time
scales for $V_0=10 E_R$ range from $10$\,ms to $100$\,ms.)  For the
other commonly used species $^6$Li, the lighter mass means shorter
time scales than for $^{40}$K.  For an optical lattice laser
wavelength of $1064\,$nm, the corresponding Gaussian relaxation time
scale is about $0.4$\,ms and $1.5$\,ms for exponential regime, with
$V_0 =5E_R$.  Hence, for optical lattice laser
strengths that are not too large,
the relaxation times are well within experimental lifetime of the cold
atom systems.  

Finally, we point out that it is not necessary to use a ring or
one-dimensional geometry (as studied
in this paper) to see the thermalisation physics we have
presented. As long as the inelastic scattering
length is small compared to the size of the bath, we expect the
qualitative aspects of thermalisation to survive, although the precise
values for the various crossovers and thresholds will change depending
on the dimensionality of the bath and the nature of subsystem-bath coupling.

\section{Conclusions}
\label{sec:conclusions}

We have presented an account of the thermalisation of a local
subsystem within a closed quantum system described by a lattice of
interacting fermions.  The subsystem thermalises in the sense that its
reduced density matrix approaches the form expected for a canonical
thermal ensemble.  This thermalisation occurs over a wide range of
system parameters for surprisingly small systems.  The equilibrium
state depends very little on the strength of subsystem-bath coupling
provided it meets the following two conditions. Most importantly, the
coupling strength needs to be large enough to mix the eigenstates of
the uncoupled system non-perturbatively. Secondly, the coupling
strength must not be so large that the boundary effects associated
with the coupling dominate the behaviour.  We also find that
small lattice clusters thermalise for a range of interaction
strengths, provided $U$ is large enough that the system is away from
integrability at $U=0$.  We were also able to demonstrate that the
energy width of the initial pure state of the bath has virtually no
effect on the subsystem state at long times.  This was found for a
range of energy widths spanning nearly two orders of magnitude.

Numerically, we demonstrated the relationship between subsystem
thermalisation at long times and the eigenstate thermalisation
hypothesis.  We further quantified the extent to which eigenstate
thermalisation holds by measuring the spread of eigenstate expectation
values for subsystem occupation probabilities. Using generic results
for the eigenvectors of perturbed quantum systems in random matrix
theory, we were able to derive theoretically a coupling-strength
threshold $\lambdaeth$ for thermalisation which is in qualitative
agreement with our numerical threshold $\lambdathermal$.  
This establishes a link between
the eigenstates of weakly-coupled bipartite quantum systems and
eigenstate thermalisation. As this result employed only random matrix
theory, our conclusions should be quite general for non-integrable
systems, provided that the system is prepared at an energy far from
the ground state where correlations may become important.

We were also able to understand the system-size scaling of the
breakdown of thermalisation seen in our numerics for interacting
fermions by considering a coupling-strength threshold, $\lambdanp$,
below which non-perturbative mixing of $\lambda=0$ eigenstates does
not occur. We demonstrated that this non-perturbative threshold
$\lambdanp$ has virtually the form as $\lambdaeth$. Moreover, 
these have the same system-size scaling as the empirical $\lambdathermal$.

We deduced that these thresholds for thermalisation should tend to
zero exponentially in the system size.  We also attribute the lack of
non-perturbative mixing as the reason for the lack of thermalisation
for the weak-interaction limit of the small systems we studied. For
very large Hubbard rings, we predict that non-perturbative mixing does
occur for any non-zero interaction $U$.

During preparation of this manuscript we became aware of unpublished
work by Neuenhahn and Marquardt~\cite{Neuenhahn2010} which also
studies eigenstate thermalisation using random-matrix-theory results
for eigenstate overlaps.  The authors study the momentum distribution
of interacting fermions on an entire closed system, in contrast to the
local observables on bipartite quantum systems considered in this
work.

\begin{acknowledgments}
  SG~acknowledges financial support from $\text{EPSRC}$ DTA funding
  and The Leverhulme Trust under grant no.~F/00114/B6. AH acknowledges
  financial support in the earlier part of the project from an
  $\text{EPSRC}$ Advanced Research Fellowship. We acknowledge
  insightful discussions with John Chalker, Fabian Essler, Stefan Kuhr
  and Michael K\"ohl.
\end{acknowledgments}

\appendix
\section{Overlaps and related distributions}
\label{sec:appendix}
We start with the distribution of overlaps $X=\inner{sb}{A}$ in
Eq.~\eqref{eq:overlapdist_collapse} with zero mean, 
variance $\expectation{X^2}=\sigma^2_{Asb}$
and fourth moment $\expectation{X^4}= (3+\gamma)\sigma^4_{Asb}$.
Since the projection $P^A_s$ is given by
$\sum_b \inner{A}{sb}\inner{sb}{A}$, we first consider the distribution,
$P^\pldos$ for $|\inner{A}{sb}|^2$:
\begin{align}
  \begin{split}
    P^\pldos&\big(|\inner{A}{sb}|^2 = Y_{Asb}\big)\\
    &= \int_{-\infty}^{\infty}\!\!\! dX_{Asb}  P(X_{Asb})
    \delta(Y_{Asb}-X_{Asb}^2) 
  \end{split}\notag\\
  &= \frac{1}{\sigma_{Asb} \sqrt{ Y_{Asb}}} \,
  F\left(\frac{\sqrt{Y_{Asb}}}{\sigma_{Asb}}\right) \label{eq:P2}
\end{align}
which has mean $\sigma^2_{Asb}$ and variance $(2+\gamma)\sigma^4_{Asb}$.
Then, the eigenstate projection values have the distribution 
$P^{\text{EP}}$, given by
\begin{multline}
  P^{\text{EP}}(P_s^A = W) =\\
  \prod_{i=1}^{\nstate_B}\left(\int_0^\infty \!\!\! dY_i  P^\pldos(Y_i)\right)
  \, 
  \delta\Big(W-\sum_{b=1}^{\nstate_B} Y_b\Big)
\end{multline}
where $\nstate_B$ is the total number of bath states and, since $A$ and $s$
are fixed, the notation is abbreviated such that $Y_{Asb}
\longrightarrow Y_b$.  Since this is a sum of many independent random
variables, albeit from different probability distributions, it is
reasonable to ask if a central limit exists.  Indeed, the Lyapunov
condition for a generalised central limit does
hold~\cite{Billingsley}.
To find this central limit, we adopt the standard procedure of
factorising the integrals in Fourier space.  Upon taking the Fourier
transform
\begin{equation}
\tilde{P}^{\text{EP}}(k) = \int_0^\infty \!\!dW e^{ikW} P^{\text{EP}}(W)
\end{equation}
the contributions from each of the bath states factorise such that
\begin{equation}
\tilde{P}^{\text{EP}}(k) = \prod_b \tilde{P}^\pldos_b(k)
\end{equation}
where
\begin{align}
\tilde{P}^\pldos_b(k) 
&= \int_0^\infty \!\!dY_b \,e^{ikY_b} \,P^\pldos(Y_b)\notag \\
&\simeq 1 +ik \sigma^2_{Asb} -\frac{3+\gamma}{2}(k \sigma^2_{Asb})^2
\end{align}
where the series has been truncated to second order in $k$.  The
logarithm of $\tilde{P}^{\text{EP}}(k)$ takes the form of the series
\begin{equation}
\log \tilde{P}^{\text{EP}}(k) \simeq\sum_b \left[ ik\sigma^2_{Asb} -
  \left(1+\frac{\gamma}{2}\right)(k\sigma^2_{Asb})^2\right]\,.
\end{equation}
The coefficient to the term linear in $k$ is simply $i\mu_\EP$
and the coefficient to the $k^2$ term is  $-\sigma^2_\EP/2$ where
$\mu_\EP$ and $\sigma_\EP$ are defined in \eqref{eq:EPstats}.
We have dropped terms of higher order of the form
$\sum_b (k\sigma_{Asb}^2)^n$. Using \eqref{eq:LDOS} and following the
same argument that leads to \eqref{eq:sigma4sum}, 
\begin{equation}
k^n \sum_b\sigma^{2n}_{A,sb} \sim \frac{k^n \mu_{\EP,As} }{ [W_Lg(E_A)]^{n-1}}
\end{equation}
where $g(E_A)$ is the density of states at $E_A$ and
$\omega_{ss}(E_A)$ is the reduced density matrix for the canonical thermal state
\eqref{eq:canonstate}. Therefore, we see that the truncation of the
series is reasonable for $k \ll g W_L \propto \lambda^2 J/\Delta$. 

Upon re-exponentiating the series, we see the bulk of the distribution
$\tilde{P}^\EP(k)$ may be described accurately with $k$ up to the scale
of $1/\sigma_\EP\propto (J/\Delta)^{1/2}$,
since the central limit only breaks down at $k\sim J/\Delta$. (This condition is
readily met in our numerics when the coupling strength $\lambda$ is
large enough for the subsystem to approach thermalisation.)
Exponentiating and inverting the Fourier transform yields the
distribution for eigenstate expectation-values:
\begin{equation}
P^{\text{EP}}(W) = \frac{1}{\sqrt{2 \pi \sigma_\EP}} 
\exp \left(-\frac{(W - \mu_\EP)^2}{2\sigma_\EP^2}\right)\,,
\end{equation}
which is a Normal distribution with mean $\mu_\EP$ and variance $\sigma_\EP$. 

\section{Coupling matrix}
\label{sec:tracevsq}
In this section, we estimate the magnitude of a matrix element
of the coupling matrix $V$ as defined in \eqref{eq:2linkH}.
As discussed in Section~\ref{sec:sizescaling}, the coupling matrix
involves only single-particle hopping between the subsystem and the
bath. So, it should connect states not further apart in energy than
the single-particle bandwidth $4J$. 

\begin{figure}[!b]
 \begin{centering}
   \includegraphics[scale=0.5]{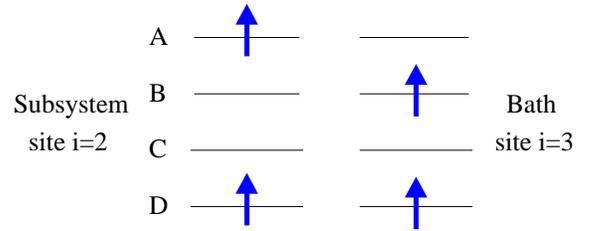}
   \caption[A diagram showing the four Fock states for the two sites
   occupied by spin-up fermions, across a coupling link.]{A diagram
     showing four possible occupations of two sites (across the
     coupling 
     link at $i$=2 and 3) by spin-up fermions, irrespective of the
     configuration of spin-down fermions on these sites.}
   \label{fockpic}
 \end{centering}
\end{figure}

We will consider the
coupling matrix to be a banded matrix where the non-zero elements form
a band of full width $2W= 4J$. While enumerating the size of individual
matrix elements is not possible without full diagonalisation of the
$\lambda=0$ Hamiltonian, the quantity $\Tr\, V^2$ is basis-independent
and may be found readily in the Fock basis, with states $\ket{F_i}$,
where particles are localised. In
this case:
$\Tr(V^2) = \sum_{ij} |\bra{F_i} V \ket{F_j}|^2 $.

The matrix $V$ does not change the total particle number.
To keep the description straightforward, we consider the case of
exactly half filling. This is demonstrated in Fig.~\ref{fockpic}.
For each basis state $\bra{F_i}$, there are at
most only four other basis states $\ket{F_j}$ which are related by
hopping a single fermion (spin up or down) between the subsystem and
the bath \emph{via} either one of the two subsystem-bath links.  As
the lattice is taken to be exactly half-filled, for each spin and for
each topological link between subsystem and bath, half of the Fock
states have a filled site adjacent to an empty site across each
coupling link. This diagram
shows four possible occupations of two sites (across the coupling
link at $i$=2 and 3) by spin-up fermions, irrespective of the
configuration of spin-down fermions on these sites.  At half filling,
the full $L$-site Fock states may be divided up into four groups
containing equal numbers of states, with each group having the spin-up
occupations A, B, C and D (as labelled in the figure).  Each state in
groups A and B can couple to one other Fock state, with matrix element
$\lambda J$, but states in groups C and D couple to no other Fock
states.  Spin-down fermions do not affect these matrix elements.

Therefore, each spin and each subsystem-bath link
contributes $\nstate J^2/2$ to the trace where $\nstate$ 
is the dimension of the
Hilbert space of the composite system.
There are contributions from
two links and two spin species.
Hence, we obtain 
\begin{equation}
\Tr(V^2) = 2 \nstate J^2\,.
\end{equation} 


\bibliography{biblio2,nanotherm,books}

\begin{thebibliography}{54}%
\makeatletter
\providecommand \@ifxundefined [1]{%
 \@ifx{#1\undefined}
}%
\providecommand \@ifnum [1]{%
 \ifnum #1\expandafter \@firstoftwo
 \else \expandafter \@secondoftwo
 \fi
}%
\providecommand \@ifx [1]{%
 \ifx #1\expandafter \@firstoftwo
 \else \expandafter \@secondoftwo
 \fi
}%
\providecommand \natexlab [1]{#1}%
\providecommand \enquote  [1]{``#1''}%
\providecommand \bibnamefont  [1]{#1}%
\providecommand \bibfnamefont [1]{#1}%
\providecommand \citenamefont [1]{#1}%
\providecommand \href@noop [0]{\@secondoftwo}%
\providecommand \href [0]{\begingroup \@sanitize@url \@href}%
\providecommand \@href[1]{\@@startlink{#1}\@@href}%
\providecommand \@@href[1]{\endgroup#1\@@endlink}%
\providecommand \@sanitize@url [0]{\catcode `\\12\catcode `\$12\catcode
  `\&12\catcode `\#12\catcode `\^12\catcode `\_12\catcode `\%12\relax}%
\providecommand \@@startlink[1]{}%
\providecommand \@@endlink[0]{}%
\providecommand \url  [0]{\begingroup\@sanitize@url \@url }%
\providecommand \@url [1]{\endgroup\@href {#1}{\urlprefix }}%
\providecommand \urlprefix  [0]{URL }%
\providecommand \Eprint [0]{\href }%
\providecommand \doibase [0]{http://dx.doi.org/}%
\providecommand \selectlanguage [0]{\@gobble}%
\providecommand \bibinfo  [0]{\@secondoftwo}%
\providecommand \bibfield  [0]{\@secondoftwo}%
\providecommand \translation [1]{[#1]}%
\providecommand \BibitemOpen [0]{}%
\providecommand \bibitemStop [0]{}%
\providecommand \bibitemNoStop [0]{.\EOS\space}%
\providecommand \EOS [0]{\spacefactor3000\relax}%
\providecommand \BibitemShut  [1]{\csname bibitem#1\endcsname}%
\let\auto@bib@innerbib\@empty
\bibitem [{\citenamefont {Cazalilla}\ and\ \citenamefont
  {Rigol}(2010)}]{Cazalilla2010}%
  \BibitemOpen
  \bibfield  {author} {\bibinfo {author} {\bibfnamefont {M.~A.}\ \bibnamefont
  {Cazalilla}}\ and\ \bibinfo {author} {\bibfnamefont {M.}~\bibnamefont
  {Rigol}},\ }\href {\doibase 10.1088/1367-2630/12/5/055006} {\bibfield
  {journal} {\bibinfo  {journal} {New Journal of Physics}\ }\textbf {\bibinfo
  {volume} {12}},\ \bibinfo {pages} {055006} (\bibinfo {year}
  {2010})}\BibitemShut {NoStop}%
\bibitem [{\citenamefont {Polkovnikov}\ \emph {et~al.}(2011)\citenamefont
  {Polkovnikov}, \citenamefont {Sengupta}, \citenamefont {Silva},\ and\
  \citenamefont {Vengalattore}}]{Polkovnikov2011}%
  \BibitemOpen
  \bibfield  {author} {\bibinfo {author} {\bibfnamefont {A.}~\bibnamefont
  {Polkovnikov}}, \bibinfo {author} {\bibfnamefont {K.}~\bibnamefont
  {Sengupta}}, \bibinfo {author} {\bibfnamefont {A.}~\bibnamefont {Silva}}, \
  and\ \bibinfo {author} {\bibfnamefont {M.}~\bibnamefont {Vengalattore}},\
  }\href {\doibase 10.1103/RevModPhys.83.863} {\bibfield  {journal} {\bibinfo
  {journal} {Rev. Mod. Phys.}\ }\textbf {\bibinfo {volume} {83}},\ \bibinfo
  {pages} {863} (\bibinfo {year} {2011})}\BibitemShut {NoStop}%
\bibitem [{\citenamefont {Yukalov}(2011)}]{Yukalov2011}%
  \BibitemOpen
  \bibfield  {author} {\bibinfo {author} {\bibfnamefont {V.~I.}\ \bibnamefont
  {Yukalov}},\ }\href@noop {} {\bibfield  {journal} {\bibinfo  {journal} {Laser
  Phys. Lett.}\ }\textbf {\bibinfo {volume} {8}},\ \bibinfo {pages} {485}
  (\bibinfo {year} {2011})}\BibitemShut {NoStop}%
\bibitem [{\citenamefont {Gemmer}\ \emph {et~al.}(2005)\citenamefont {Gemmer},
  \citenamefont {Michel},\ and\ \citenamefont
  {Mahler}}]{QuantumThermodynamics}%
  \BibitemOpen
  \bibfield  {author} {\bibinfo {author} {\bibfnamefont {J.}~\bibnamefont
  {Gemmer}}, \bibinfo {author} {\bibfnamefont {M.}~\bibnamefont {Michel}}, \
  and\ \bibinfo {author} {\bibfnamefont {G.}~\bibnamefont {Mahler}},\
  }\href@noop {} {\emph {\bibinfo {title} {Quantum Thermodynamics: Emergence of
  Thermodynamic Behavior Within Composite Quantum Systems}}},\ Lecture Notes in
  Physics\ (\bibinfo  {publisher} {Springer},\ \bibinfo {year}
  {2005})\BibitemShut {NoStop}%
\bibitem [{\citenamefont {Saito}\ \emph {et~al.}(1996)\citenamefont {Saito},
  \citenamefont {Takesue},\ and\ \citenamefont {Miyashita}}]{Saito1996}%
  \BibitemOpen
  \bibfield  {author} {\bibinfo {author} {\bibfnamefont {K.}~\bibnamefont
  {Saito}}, \bibinfo {author} {\bibfnamefont {S.}~\bibnamefont {Takesue}}, \
  and\ \bibinfo {author} {\bibfnamefont {S.}~\bibnamefont {Miyashita}},\ }\href
  {http://cat.inist.fr/?aModele=afficheN\&amp;cpsidt=3102680} {\bibfield
  {journal} {\bibinfo  {journal} {Journal of the Physical Society of Japan}\
  }\textbf {\bibinfo {volume} {65}},\ \bibinfo {pages} {1243} (\bibinfo {year}
  {1996})}\BibitemShut {NoStop}%
\bibitem [{\citenamefont {Yuan}\ \emph {et~al.}(2009)\citenamefont {Yuan},
  \citenamefont {Katsnelson},\ and\ \citenamefont {{De Raedt}}}]{Yuan2009}%
  \BibitemOpen
  \bibfield  {author} {\bibinfo {author} {\bibfnamefont {S.}~\bibnamefont
  {Yuan}}, \bibinfo {author} {\bibfnamefont {M.~I.}\ \bibnamefont
  {Katsnelson}}, \ and\ \bibinfo {author} {\bibfnamefont {H.}~\bibnamefont {{De
  Raedt}}},\ }\href {\doibase 10.1143/JPSJ.78.094003} {\bibfield  {journal}
  {\bibinfo  {journal} {Journal of the Physical Society of Japan}\ }\textbf
  {\bibinfo {volume} {78}},\ \bibinfo {pages} {094003} (\bibinfo {year}
  {2009})}\BibitemShut {NoStop}%
\bibitem [{\citenamefont {Jin}\ \emph {et~al.}(2010)\citenamefont {Jin},
  \citenamefont {{De Raedt}}, \citenamefont {Yuan}, \citenamefont {Katsnelson},
  \citenamefont {Miyashita},\ and\ \citenamefont {Michielsen}}]{Jin2010}%
  \BibitemOpen
  \bibfield  {author} {\bibinfo {author} {\bibfnamefont {F.}~\bibnamefont
  {Jin}}, \bibinfo {author} {\bibfnamefont {H.}~\bibnamefont {{De Raedt}}},
  \bibinfo {author} {\bibfnamefont {S.}~\bibnamefont {Yuan}}, \bibinfo {author}
  {\bibfnamefont {M.~I.}\ \bibnamefont {Katsnelson}}, \bibinfo {author}
  {\bibfnamefont {S.}~\bibnamefont {Miyashita}}, \ and\ \bibinfo {author}
  {\bibfnamefont {K.}~\bibnamefont {Michielsen}},\ }\href {\doibase
  10.1143/JPSJ.79.124005} {\bibfield  {journal} {\bibinfo  {journal} {Journal
  of the Physical Society of Japan}\ }\textbf {\bibinfo {volume} {79}},\
  \bibinfo {pages} {124005} (\bibinfo {year} {2010})}\BibitemShut {NoStop}%
\bibitem [{\citenamefont {Henrich}\ \emph {et~al.}(2005)\citenamefont
  {Henrich}, \citenamefont {Michel}, \citenamefont {Hartmann}, \citenamefont
  {Mahler},\ and\ \citenamefont {Gemmer}}]{Henrich2005}%
  \BibitemOpen
  \bibfield  {author} {\bibinfo {author} {\bibfnamefont {M.}~\bibnamefont
  {Henrich}}, \bibinfo {author} {\bibfnamefont {M.}~\bibnamefont {Michel}},
  \bibinfo {author} {\bibfnamefont {M.}~\bibnamefont {Hartmann}}, \bibinfo
  {author} {\bibfnamefont {G.}~\bibnamefont {Mahler}}, \ and\ \bibinfo {author}
  {\bibfnamefont {J.}~\bibnamefont {Gemmer}},\ }\href {\doibase
  10.1103/PhysRevE.72.026104} {\bibfield  {journal} {\bibinfo  {journal}
  {Physical Review E}\ }\textbf {\bibinfo {volume} {72}},\ \bibinfo {pages}
  {026104} (\bibinfo {year} {2005})}\BibitemShut {NoStop}%
\bibitem [{\citenamefont {Gogolin}\ \emph {et~al.}(2011)\citenamefont
  {Gogolin}, \citenamefont {M\"{u}ller},\ and\ \citenamefont
  {Eisert}}]{Gogolin2011}%
  \BibitemOpen
  \bibfield  {author} {\bibinfo {author} {\bibfnamefont {C.}~\bibnamefont
  {Gogolin}}, \bibinfo {author} {\bibfnamefont {M.}~\bibnamefont {M\"{u}ller}},
  \ and\ \bibinfo {author} {\bibfnamefont {J.}~\bibnamefont {Eisert}},\ }\href
  {\doibase 10.1103/PhysRevLett.106.040401} {\bibfield  {journal} {\bibinfo
  {journal} {Physical Review Letters}\ }\textbf {\bibinfo {volume} {106}},\
  \bibinfo {pages} {040401} (\bibinfo {year} {2011})}\BibitemShut {NoStop}%
\bibitem [{\citenamefont {Cho}\ and\ \citenamefont {Kim}(2010)}]{Cho2010}%
  \BibitemOpen
  \bibfield  {author} {\bibinfo {author} {\bibfnamefont {J.}~\bibnamefont
  {Cho}}\ and\ \bibinfo {author} {\bibfnamefont {M.~S.}\ \bibnamefont {Kim}},\
  }\href {\doibase 10.1103/PhysRevLett.104.170402} {\bibfield  {journal}
  {\bibinfo  {journal} {Physical Review Letters}\ }\textbf {\bibinfo {volume}
  {104}},\ \bibinfo {pages} {170402} (\bibinfo {year} {2010})}\BibitemShut
  {NoStop}%
\bibitem [{\citenamefont {Tasaki}(1998)}]{Tasaki1998}%
  \BibitemOpen
  \bibfield  {author} {\bibinfo {author} {\bibfnamefont {H.}~\bibnamefont
  {Tasaki}},\ }\href@noop {} {\bibfield  {journal} {\bibinfo  {journal}
  {Physical Review Letters}\ ,\ \bibinfo {pages} {1373}} (\bibinfo {year}
  {1998})}\BibitemShut {NoStop}%
\bibitem [{\citenamefont {Goldstein}\ \emph {et~al.}(2006)\citenamefont
  {Goldstein}, \citenamefont {Lebowitz}, \citenamefont {Tumulka},\ and\
  \citenamefont {Zangh\`{\i}}}]{Goldstein2006}%
  \BibitemOpen
  \bibfield  {author} {\bibinfo {author} {\bibfnamefont {S.}~\bibnamefont
  {Goldstein}}, \bibinfo {author} {\bibfnamefont {J.}~\bibnamefont {Lebowitz}},
  \bibinfo {author} {\bibfnamefont {R.}~\bibnamefont {Tumulka}}, \ and\
  \bibinfo {author} {\bibfnamefont {N.}~\bibnamefont {Zangh\`{\i}}},\ }\href
  {\doibase 10.1103/PhysRevLett.96.050403} {\bibfield  {journal} {\bibinfo
  {journal} {Physical Review Letters}\ }\textbf {\bibinfo {volume} {96}},\
  \bibinfo {pages} {050403} (\bibinfo {year} {2006})}\BibitemShut {NoStop}%
\bibitem [{\citenamefont {Popescu}\ \emph {et~al.}(2006)\citenamefont
  {Popescu}, \citenamefont {Short},\ and\ \citenamefont
  {Winter}}]{Popescu2006}%
  \BibitemOpen
  \bibfield  {author} {\bibinfo {author} {\bibfnamefont {S.}~\bibnamefont
  {Popescu}}, \bibinfo {author} {\bibfnamefont {A.~J.}\ \bibnamefont {Short}},
  \ and\ \bibinfo {author} {\bibfnamefont {A.}~\bibnamefont {Winter}},\ }\href
  {\doibase 10.1038/nphys444} {\bibfield  {journal} {\bibinfo  {journal}
  {Nature Physics}\ }\textbf {\bibinfo {volume} {2}},\ \bibinfo {pages} {754}
  (\bibinfo {year} {2006})}\BibitemShut {NoStop}%
\bibitem [{\citenamefont {Reimann}(2007)}]{Reimann2007}%
  \BibitemOpen
  \bibfield  {author} {\bibinfo {author} {\bibfnamefont {P.}~\bibnamefont
  {Reimann}},\ }\href {\doibase 10.1103/PhysRevLett.99.160404} {\bibfield
  {journal} {\bibinfo  {journal} {Physical Review Letters}\ }\textbf {\bibinfo
  {volume} {99}},\ \bibinfo {pages} {160404} (\bibinfo {year}
  {2007})}\BibitemShut {NoStop}%
\bibitem [{\citenamefont {Reimann}(2008)}]{Reimann2008}%
  \BibitemOpen
  \bibfield  {author} {\bibinfo {author} {\bibfnamefont {P.}~\bibnamefont
  {Reimann}},\ }\href {\doibase 10.1103/PhysRevLett.101.190403} {\bibfield
  {journal} {\bibinfo  {journal} {Physical Review Letters}\ }\textbf {\bibinfo
  {volume} {101}},\ \bibinfo {pages} {190403} (\bibinfo {year}
  {2008})}\BibitemShut {NoStop}%
\bibitem [{\citenamefont {Linden}\ \emph {et~al.}(2009)\citenamefont {Linden},
  \citenamefont {Popescu}, \citenamefont {Short},\ and\ \citenamefont
  {Winter}}]{Linden2009}%
  \BibitemOpen
  \bibfield  {author} {\bibinfo {author} {\bibfnamefont {N.}~\bibnamefont
  {Linden}}, \bibinfo {author} {\bibfnamefont {S.}~\bibnamefont {Popescu}},
  \bibinfo {author} {\bibfnamefont {A.}~\bibnamefont {Short}}, \ and\ \bibinfo
  {author} {\bibfnamefont {A.}~\bibnamefont {Winter}},\ }\href {\doibase
  10.1103/PhysRevE.79.061103} {\bibfield  {journal} {\bibinfo  {journal}
  {Physical Review E}\ }\textbf {\bibinfo {volume} {79}},\ \bibinfo {pages}
  {061103} (\bibinfo {year} {2009})}\BibitemShut {NoStop}%
\bibitem [{\citenamefont {Gogolin}(2010)}]{Gogolin2010}%
  \BibitemOpen
  \bibfield  {author} {\bibinfo {author} {\bibfnamefont {C.}~\bibnamefont
  {Gogolin}},\ }\href {\doibase 10.1103/PhysRevE.81.051127} {\bibfield
  {journal} {\bibinfo  {journal} {Physical Review E}\ }\textbf {\bibinfo
  {volume} {81}},\ \bibinfo {pages} {051127} (\bibinfo {year}
  {2010})}\BibitemShut {NoStop}%
\bibitem [{\citenamefont {Lychkovskiy}(2010)}]{Lychkovskiy2010}%
  \BibitemOpen
  \bibfield  {author} {\bibinfo {author} {\bibfnamefont {O.}~\bibnamefont
  {Lychkovskiy}},\ }\href {\doibase 10.1103/PhysRevE.82.011123} {\bibfield
  {journal} {\bibinfo  {journal} {Physical Review E}\ }\textbf {\bibinfo
  {volume} {82}},\ \bibinfo {pages} {011123} (\bibinfo {year}
  {2010})}\BibitemShut {NoStop}%
\bibitem [{\citenamefont {Deutsch}(1991)}]{Deutsch1991}%
  \BibitemOpen
  \bibfield  {author} {\bibinfo {author} {\bibfnamefont {J.~M.}\ \bibnamefont
  {Deutsch}},\ }\href@noop {} {\bibfield  {journal} {\bibinfo  {journal}
  {Physical Review A}\ }\textbf {\bibinfo {volume} {43}},\ \bibinfo {pages}
  {2046} (\bibinfo {year} {1991})}\BibitemShut {NoStop}%
\bibitem [{\citenamefont {Srednicki}(1994)}]{Srednicki1994}%
  \BibitemOpen
  \bibfield  {author} {\bibinfo {author} {\bibfnamefont {M.}~\bibnamefont
  {Srednicki}},\ }\href@noop {} {\bibfield  {journal} {\bibinfo  {journal}
  {Physical Review E}\ }\textbf {\bibinfo {volume} {50}},\ \bibinfo {pages}
  {888} (\bibinfo {year} {1994})}\BibitemShut {NoStop}%
\bibitem [{\citenamefont {Rigol}\ \emph {et~al.}(2008)\citenamefont {Rigol},
  \citenamefont {Dunjko},\ and\ \citenamefont {Olshanii}}]{Rigol2008}%
  \BibitemOpen
  \bibfield  {author} {\bibinfo {author} {\bibfnamefont {M.}~\bibnamefont
  {Rigol}}, \bibinfo {author} {\bibfnamefont {V.}~\bibnamefont {Dunjko}}, \
  and\ \bibinfo {author} {\bibfnamefont {M.}~\bibnamefont {Olshanii}},\ }\href
  {\doibase 10.1038/nature06838} {\bibfield  {journal} {\bibinfo  {journal}
  {Nature}\ }\textbf {\bibinfo {volume} {452}},\ \bibinfo {pages} {854}
  (\bibinfo {year} {2008})}\BibitemShut {NoStop}%
\bibitem [{\citenamefont {Sherson}\ \emph {et~al.}(2010)\citenamefont
  {Sherson}, \citenamefont {Weitenberg}, \citenamefont {Endres}, \citenamefont
  {Cheneau}, \citenamefont {Bloch},\ and\ \citenamefont {Kuhr}}]{Sherson2010}%
  \BibitemOpen
  \bibfield  {author} {\bibinfo {author} {\bibfnamefont {J.~F.}\ \bibnamefont
  {Sherson}}, \bibinfo {author} {\bibfnamefont {C.}~\bibnamefont {Weitenberg}},
  \bibinfo {author} {\bibfnamefont {M.}~\bibnamefont {Endres}}, \bibinfo
  {author} {\bibfnamefont {M.}~\bibnamefont {Cheneau}}, \bibinfo {author}
  {\bibfnamefont {I.}~\bibnamefont {Bloch}}, \ and\ \bibinfo {author}
  {\bibfnamefont {S.}~\bibnamefont {Kuhr}},\ }\href {\doibase
  10.1038/nature09378} {\bibfield  {journal} {\bibinfo  {journal} {Nature}\
  }\textbf {\bibinfo {volume} {467}},\ \bibinfo {pages} {68} (\bibinfo {year}
  {2010})}\BibitemShut {NoStop}%
\bibitem [{\citenamefont {Bakr}\ \emph {et~al.}(2010)\citenamefont {Bakr},
  \citenamefont {Peng}, \citenamefont {Tai}, \citenamefont {Ma}, \citenamefont
  {Simon}, \citenamefont {Gillen}, \citenamefont {Fölling}, \citenamefont
  {Pollet},\ and\ \citenamefont {Greiner}}]{Bakr2010}%
  \BibitemOpen
  \bibfield  {author} {\bibinfo {author} {\bibfnamefont {W.~S.}\ \bibnamefont
  {Bakr}}, \bibinfo {author} {\bibfnamefont {A.}~\bibnamefont {Peng}}, \bibinfo
  {author} {\bibfnamefont {M.~E.}\ \bibnamefont {Tai}}, \bibinfo {author}
  {\bibfnamefont {R.}~\bibnamefont {Ma}}, \bibinfo {author} {\bibfnamefont
  {J.}~\bibnamefont {Simon}}, \bibinfo {author} {\bibfnamefont {J.~I.}\
  \bibnamefont {Gillen}}, \bibinfo {author} {\bibfnamefont {S.}~\bibnamefont
  {Fölling}}, \bibinfo {author} {\bibfnamefont {L.}~\bibnamefont {Pollet}}, \
  and\ \bibinfo {author} {\bibfnamefont {M.}~\bibnamefont {Greiner}},\ }\href
  {\doibase 10.1126/science.1192368} {\bibfield  {journal} {\bibinfo  {journal}
  {Science}\ }\textbf {\bibinfo {volume} {329}},\ \bibinfo {pages} {547}
  (\bibinfo {year} {2010})}\BibitemShut {NoStop}%
\bibitem [{\citenamefont {Rigol}\ \emph {et~al.}(2007)\citenamefont {Rigol},
  \citenamefont {Dunjko}, \citenamefont {Yurovsky},\ and\ \citenamefont
  {Olshanii}}]{Rigol2007}%
  \BibitemOpen
  \bibfield  {author} {\bibinfo {author} {\bibfnamefont {M.}~\bibnamefont
  {Rigol}}, \bibinfo {author} {\bibfnamefont {V.}~\bibnamefont {Dunjko}},
  \bibinfo {author} {\bibfnamefont {V.}~\bibnamefont {Yurovsky}}, \ and\
  \bibinfo {author} {\bibfnamefont {M.}~\bibnamefont {Olshanii}},\ }\href
  {\doibase 10.1103/PhysRevLett.98.050405} {\bibfield  {journal} {\bibinfo
  {journal} {Physical Review Letters}\ }\textbf {\bibinfo {volume} {98}},\
  \bibinfo {pages} {050405} (\bibinfo {year} {2007})}\BibitemShut {NoStop}%
\bibitem [{\citenamefont {Berry}(1977)}]{Berry1977}%
  \BibitemOpen
  \bibfield  {author} {\bibinfo {author} {\bibfnamefont {M.~V.}\ \bibnamefont
  {Berry}},\ }\href {http://stacks.iop.org/0305-4470/10/i=12/a=016} {\bibfield
  {journal} {\bibinfo  {journal} {Journal of Physics A: Mathematical and
  General}\ }\textbf {\bibinfo {volume} {10}},\ \bibinfo {pages} {2083}
  (\bibinfo {year} {1977})}\BibitemShut {NoStop}%
\bibitem [{\citenamefont {Rigol}\ and\ \citenamefont
  {Santos}(2010)}]{Rigol2010}%
  \BibitemOpen
  \bibfield  {author} {\bibinfo {author} {\bibfnamefont {M.}~\bibnamefont
  {Rigol}}\ and\ \bibinfo {author} {\bibfnamefont {L.}~\bibnamefont {Santos}},\
  }\href {\doibase 10.1103/PhysRevA.82.011604} {\bibfield  {journal} {\bibinfo
  {journal} {Physical Review A}\ }\textbf {\bibinfo {volume} {82}},\ \bibinfo
  {pages} {011604} (\bibinfo {year} {2010})}\BibitemShut {NoStop}%
\bibitem [{\citenamefont {Ji}\ and\ \citenamefont {Fine}(2011)}]{Ji2011}%
  \BibitemOpen
  \bibfield  {author} {\bibinfo {author} {\bibfnamefont {K.}~\bibnamefont
  {Ji}}\ and\ \bibinfo {author} {\bibfnamefont {B.}~\bibnamefont {Fine}},\
  }\href {\doibase 10.1103/PhysRevLett.107.050401} {\bibfield  {journal}
  {\bibinfo  {journal} {Physical Review Letters}\ }\textbf {\bibinfo {volume}
  {107}},\ \bibinfo {pages} {050401} (\bibinfo {year} {2011})}\BibitemShut
  {NoStop}%
\bibitem [{\citenamefont {Fine}(2009)}]{Fine2009}%
  \BibitemOpen
  \bibfield  {author} {\bibinfo {author} {\bibfnamefont {B.~V.}\ \bibnamefont
  {Fine}},\ }\href {\doibase 10.1103/PhysRevE.80.051130} {\bibfield  {journal}
  {\bibinfo  {journal} {Physical Review E}\ }\textbf {\bibinfo {volume} {80}},\
  \bibinfo {pages} {051130} (\bibinfo {year} {2009})}\BibitemShut {NoStop}%
\bibitem [{\citenamefont {Canovi}\ \emph {et~al.}(2011)\citenamefont {Canovi},
  \citenamefont {Rossini}, \citenamefont {Fazio}, \citenamefont {Santoro},\
  and\ \citenamefont {Silva}}]{Canovi2011}%
  \BibitemOpen
  \bibfield  {author} {\bibinfo {author} {\bibfnamefont {E.}~\bibnamefont
  {Canovi}}, \bibinfo {author} {\bibfnamefont {D.}~\bibnamefont {Rossini}},
  \bibinfo {author} {\bibfnamefont {R.}~\bibnamefont {Fazio}}, \bibinfo
  {author} {\bibfnamefont {G.~E.}\ \bibnamefont {Santoro}}, \ and\ \bibinfo
  {author} {\bibfnamefont {A.}~\bibnamefont {Silva}},\ }\href {\doibase
  10.1103/PhysRevB.83.094431} {\bibfield  {journal} {\bibinfo  {journal} {Phys.
  Rev. B}\ }\textbf {\bibinfo {volume} {83}},\ \bibinfo {pages} {094431}
  (\bibinfo {year} {2011})}\BibitemShut {NoStop}%
\bibitem [{\citenamefont {Lesanovsky}\ \emph {et~al.}(2010)\citenamefont
  {Lesanovsky}, \citenamefont {Olmos},\ and\ \citenamefont
  {Garrahan}}]{Lesanovsky2010}%
  \BibitemOpen
  \bibfield  {author} {\bibinfo {author} {\bibfnamefont {I.}~\bibnamefont
  {Lesanovsky}}, \bibinfo {author} {\bibfnamefont {B.}~\bibnamefont {Olmos}}, \
  and\ \bibinfo {author} {\bibfnamefont {J.}~\bibnamefont {Garrahan}},\ }\href
  {\doibase 10.1103/PhysRevLett.105.100603} {\bibfield  {journal} {\bibinfo
  {journal} {Physical Review Letters}\ }\textbf {\bibinfo {volume} {105}},\
  \bibinfo {pages} {100603} (\bibinfo {year} {2010})}\BibitemShut {NoStop}%
\bibitem [{\citenamefont {Ates}\ \emph {et~al.}(2012)\citenamefont {Ates},
  \citenamefont {Garrahan},\ and\ \citenamefont {Lesanovsky}}]{Ates2012}%
  \BibitemOpen
  \bibfield  {author} {\bibinfo {author} {\bibfnamefont {C.}~\bibnamefont
  {Ates}}, \bibinfo {author} {\bibfnamefont {J.}~\bibnamefont {Garrahan}}, \
  and\ \bibinfo {author} {\bibfnamefont {I.}~\bibnamefont {Lesanovsky}},\
  }\href {\doibase 10.1103/PhysRevLett.108.110603} {\bibfield  {journal}
  {\bibinfo  {journal} {Physical Review Letters}\ }\textbf {\bibinfo {volume}
  {108}},\ \bibinfo {pages} {110603} (\bibinfo {year} {2012})}\BibitemShut
  {NoStop}%
\bibitem [{\citenamefont {Kinoshita}\ \emph {et~al.}(2006)\citenamefont
  {Kinoshita}, \citenamefont {Wenger},\ and\ \citenamefont
  {Weiss}}]{Kinoshita2006}%
  \BibitemOpen
  \bibfield  {author} {\bibinfo {author} {\bibfnamefont {T.}~\bibnamefont
  {Kinoshita}}, \bibinfo {author} {\bibfnamefont {T.}~\bibnamefont {Wenger}}, \
  and\ \bibinfo {author} {\bibfnamefont {D.~S.}\ \bibnamefont {Weiss}},\ }\href
  {\doibase 10.1038/nature04693} {\bibfield  {journal} {\bibinfo  {journal}
  {Nature}\ }\textbf {\bibinfo {volume} {440}},\ \bibinfo {pages} {900}
  (\bibinfo {year} {2006})}\BibitemShut {NoStop}%
\bibitem [{\citenamefont {Eckstein}\ and\ \citenamefont
  {Kollar}(2008)}]{Eckstein2008}%
  \BibitemOpen
  \bibfield  {author} {\bibinfo {author} {\bibfnamefont {M.}~\bibnamefont
  {Eckstein}}\ and\ \bibinfo {author} {\bibfnamefont {M.}~\bibnamefont
  {Kollar}},\ }\href {\doibase 10.1103/PhysRevLett.100.120404} {\bibfield
  {journal} {\bibinfo  {journal} {Phys. Rev. Lett.}\ }\textbf {\bibinfo
  {volume} {100}},\ \bibinfo {pages} {120404} (\bibinfo {year}
  {2008})}\BibitemShut {NoStop}%
\bibitem [{\citenamefont {Sotiriadis}\ \emph {et~al.}(2009)\citenamefont
  {Sotiriadis}, \citenamefont {Calabrese},\ and\ \citenamefont
  {Cardy}}]{Sotiriadis2009}%
  \BibitemOpen
  \bibfield  {author} {\bibinfo {author} {\bibfnamefont {S.}~\bibnamefont
  {Sotiriadis}}, \bibinfo {author} {\bibfnamefont {P.}~\bibnamefont
  {Calabrese}}, \ and\ \bibinfo {author} {\bibfnamefont {J.}~\bibnamefont
  {Cardy}},\ }\href {http://stacks.iop.org/0295-5075/87/i=2/a=20002} {\bibfield
   {journal} {\bibinfo  {journal} {EPL (Europhysics Letters)}\ }\textbf
  {\bibinfo {volume} {87}},\ \bibinfo {pages} {20002} (\bibinfo {year}
  {2009})}\BibitemShut {NoStop}%
\bibitem [{\citenamefont {Calabrese}\ \emph {et~al.}(2011)\citenamefont
  {Calabrese}, \citenamefont {Essler},\ and\ \citenamefont
  {Fagotti}}]{Calabrese2011}%
  \BibitemOpen
  \bibfield  {author} {\bibinfo {author} {\bibfnamefont {P.}~\bibnamefont
  {Calabrese}}, \bibinfo {author} {\bibfnamefont {F.~H.~L.}\ \bibnamefont
  {Essler}}, \ and\ \bibinfo {author} {\bibfnamefont {M.}~\bibnamefont
  {Fagotti}},\ }\href {\doibase 10.1103/PhysRevLett.106.227203} {\bibfield
  {journal} {\bibinfo  {journal} {Phys. Rev. Lett.}\ }\textbf {\bibinfo
  {volume} {106}},\ \bibinfo {pages} {227203} (\bibinfo {year}
  {2011})}\BibitemShut {NoStop}%
\bibitem [{\citenamefont {Cassidy}\ \emph {et~al.}(2011)\citenamefont
  {Cassidy}, \citenamefont {Clark},\ and\ \citenamefont {Rigol}}]{Cassidy2011}%
  \BibitemOpen
  \bibfield  {author} {\bibinfo {author} {\bibfnamefont {A.}~\bibnamefont
  {Cassidy}}, \bibinfo {author} {\bibfnamefont {C.}~\bibnamefont {Clark}}, \
  and\ \bibinfo {author} {\bibfnamefont {M.}~\bibnamefont {Rigol}},\ }\href
  {\doibase 10.1103/PhysRevLett.106.140405} {\bibfield  {journal} {\bibinfo
  {journal} {Physical Review Letters}\ }\textbf {\bibinfo {volume} {106}},\
  \bibinfo {pages} {140405} (\bibinfo {year} {2011})}\BibitemShut {NoStop}%
\bibitem [{\citenamefont {Poilblanc}(2011)}]{Poilblanc2011}%
  \BibitemOpen
  \bibfield  {author} {\bibinfo {author} {\bibfnamefont {D.}~\bibnamefont
  {Poilblanc}},\ }\href {\doibase 10.1103/PhysRevB.84.045120} {\bibfield
  {journal} {\bibinfo  {journal} {Phys. Rev. B}\ }\textbf {\bibinfo {volume}
  {84}},\ \bibinfo {pages} {045120} (\bibinfo {year} {2011})}\BibitemShut
  {NoStop}%
\bibitem [{\citenamefont {Jensen}\ and\ \citenamefont
  {Shankar}(1985)}]{Jensen1985}%
  \BibitemOpen
  \bibfield  {author} {\bibinfo {author} {\bibfnamefont {R.}~\bibnamefont
  {Jensen}}\ and\ \bibinfo {author} {\bibfnamefont {R.}~\bibnamefont
  {Shankar}},\ }\href@noop {} {\bibfield  {journal} {\bibinfo  {journal}
  {Physical Review Letters}\ }\textbf {\bibinfo {volume} {54}},\ \bibinfo
  {pages} {1879} (\bibinfo {year} {1985})}\BibitemShut {NoStop}%
\bibitem [{\citenamefont {Eckstein}\ \emph {et~al.}(2009)\citenamefont
  {Eckstein}, \citenamefont {Kollar},\ and\ \citenamefont
  {Werner}}]{Eckstein2009}%
  \BibitemOpen
  \bibfield  {author} {\bibinfo {author} {\bibfnamefont {M.}~\bibnamefont
  {Eckstein}}, \bibinfo {author} {\bibfnamefont {M.}~\bibnamefont {Kollar}}, \
  and\ \bibinfo {author} {\bibfnamefont {P.}~\bibnamefont {Werner}},\ }\href
  {\doibase 10.1103/PhysRevLett.103.056403} {\bibfield  {journal} {\bibinfo
  {journal} {Phys. Rev. Lett.}\ }\textbf {\bibinfo {volume} {103}},\ \bibinfo
  {pages} {056403} (\bibinfo {year} {2009})}\BibitemShut {NoStop}%
\bibitem [{\citenamefont {Genway}\ \emph {et~al.}(2010)\citenamefont {Genway},
  \citenamefont {Ho},\ and\ \citenamefont {Lee}}]{Genway2010}%
  \BibitemOpen
  \bibfield  {author} {\bibinfo {author} {\bibfnamefont {S.}~\bibnamefont
  {Genway}}, \bibinfo {author} {\bibfnamefont {A.}~\bibnamefont {Ho}}, \ and\
  \bibinfo {author} {\bibfnamefont {D.~K.~K.}\ \bibnamefont {Lee}},\ }\href
  {\doibase 10.1103/PhysRevLett.105.260402} {\bibfield  {journal} {\bibinfo
  {journal} {Physical Review Letters}\ }\textbf {\bibinfo {volume} {105}},\
  \bibinfo {pages} {260402} (\bibinfo {year} {2010})}\BibitemShut {NoStop}%
\bibitem [{\citenamefont {{Kollar}}\ \emph {et~al.}(2011)\citenamefont
  {{Kollar}}, \citenamefont {{Wolf}},\ and\ \citenamefont
  {{Eckstein}}}]{Kollar2011}%
  \BibitemOpen
  \bibfield  {author} {\bibinfo {author} {\bibfnamefont {M.}~\bibnamefont
  {{Kollar}}}, \bibinfo {author} {\bibfnamefont {F.~A.}\ \bibnamefont
  {{Wolf}}}, \ and\ \bibinfo {author} {\bibfnamefont {M.}~\bibnamefont
  {{Eckstein}}},\ }\href {\doibase 10.1103/PhysRevB.84.054304} {\bibfield
  {journal} {\bibinfo  {journal} {\prb}\ }\textbf {\bibinfo {volume} {84}},\
  \bibinfo {eid} {054304} (\bibinfo {year} {2011})},\ \Eprint
  {http://arxiv.org/abs/1102.2117} {arXiv:1102.2117 [cond-mat.str-el]}
  \BibitemShut {NoStop}%
\bibitem [{\citenamefont {Kottos}\ and\ \citenamefont
  {Cohen}(2001)}]{Kottos2001}%
  \BibitemOpen
  \bibfield  {author} {\bibinfo {author} {\bibfnamefont {T.}~\bibnamefont
  {Kottos}}\ and\ \bibinfo {author} {\bibfnamefont {D.}~\bibnamefont {Cohen}},\
  }\href {\doibase 10.1103/PhysRevE.64.065202} {\bibfield  {journal} {\bibinfo
  {journal} {Physical Review E}\ }\textbf {\bibinfo {volume} {64}},\ \bibinfo
  {pages} {065202} (\bibinfo {year} {2001})}\BibitemShut {NoStop}%
\bibitem [{\citenamefont {Cohen}\ \emph {et~al.}(2000)\citenamefont {Cohen},
  \citenamefont {Izrailev},\ and\ \citenamefont {Kottos}}]{Cohen2000}%
  \BibitemOpen
  \bibfield  {author} {\bibinfo {author} {\bibfnamefont {D.}~\bibnamefont
  {Cohen}}, \bibinfo {author} {\bibfnamefont {F.~M.}\ \bibnamefont {Izrailev}},
  \ and\ \bibinfo {author} {\bibfnamefont {T.}~\bibnamefont {Kottos}},\ }\href
  {http://www.ncbi.nlm.nih.gov/pubmed/11017207} {\bibfield  {journal} {\bibinfo
   {journal} {Physical review letters}\ }\textbf {\bibinfo {volume} {84}},\
  \bibinfo {pages} {2052} (\bibinfo {year} {2000})}\BibitemShut {NoStop}%
\bibitem [{\citenamefont {Hiller}\ \emph {et~al.}(2006)\citenamefont {Hiller},
  \citenamefont {Kottos},\ and\ \citenamefont {Geisel}}]{Hiller2006}%
  \BibitemOpen
  \bibfield  {author} {\bibinfo {author} {\bibfnamefont {M.}~\bibnamefont
  {Hiller}}, \bibinfo {author} {\bibfnamefont {T.}~\bibnamefont {Kottos}}, \
  and\ \bibinfo {author} {\bibfnamefont {T.}~\bibnamefont {Geisel}},\ }\href
  {\doibase 10.1103/PhysRevA.73.061604} {\bibfield  {journal} {\bibinfo
  {journal} {Physical Review A}\ }\textbf {\bibinfo {volume} {73}},\ \bibinfo
  {pages} {061604} (\bibinfo {year} {2006})}\BibitemShut {NoStop}%
\bibitem [{\citenamefont {Wigner}(1955)}]{Wigner1955}%
  \BibitemOpen
  \bibfield  {author} {\bibinfo {author} {\bibfnamefont {E.~P.}\ \bibnamefont
  {Wigner}},\ }\href@noop {} {\bibfield  {journal} {\bibinfo  {journal} {Annals
  of Mathematics}\ }\textbf {\bibinfo {volume} {62}},\ \bibinfo {pages} {548}
  (\bibinfo {year} {1955})}\BibitemShut {NoStop}%
\bibitem [{\citenamefont {Wigner}(1957)}]{Wigner1957}%
  \BibitemOpen
  \bibfield  {author} {\bibinfo {author} {\bibfnamefont {E.~P.}\ \bibnamefont
  {Wigner}},\ }\href@noop {} {\bibfield  {journal} {\bibinfo  {journal} {Annals
  of Mathematics}\ }\textbf {\bibinfo {volume} {67}},\ \bibinfo {pages} {325}
  (\bibinfo {year} {1957})}\BibitemShut {NoStop}%
\bibitem [{Note1()}]{Note1}%
  \BibitemOpen
  \bibinfo {note} {In preparation.}\BibitemShut {Stop}%
\bibitem [{\citenamefont {Feynman}(1998)}]{Feynman}%
  \BibitemOpen
  \bibfield  {author} {\bibinfo {author} {\bibfnamefont {R.}~\bibnamefont
  {Feynman}},\ }\href {http://books.google.com/books?id=Ou4ltPYiXPgC} {\emph
  {\bibinfo {title} {Statistical mechanics: a set of lectures}}},\ Advanced
  book classics\ (\bibinfo  {publisher} {Westview Press},\ \bibinfo {year}
  {1998})\BibitemShut {NoStop}%
\bibitem [{\citenamefont {Bloch}(2005)}]{Bloch2005}%
  \BibitemOpen
  \bibfield  {author} {\bibinfo {author} {\bibfnamefont {I.}~\bibnamefont
  {Bloch}},\ }\href
  {http://www.nature.com/nphys/journal/v1/n1/abs/nphys138.html} {\bibfield
  {journal} {\bibinfo  {journal} {Nature Physics}\ }\textbf {\bibinfo {volume}
  {1}},\ \bibinfo {pages} {23} (\bibinfo {year} {2005})}\BibitemShut {NoStop}%
\bibitem [{\citenamefont {Bakr}\ \emph {et~al.}(2009)\citenamefont {Bakr},
  \citenamefont {Gillen}, \citenamefont {Peng}, \citenamefont {F\"{o}lling},\
  and\ \citenamefont {Greiner}}]{Bakr2009}%
  \BibitemOpen
  \bibfield  {author} {\bibinfo {author} {\bibfnamefont {W.~S.}\ \bibnamefont
  {Bakr}}, \bibinfo {author} {\bibfnamefont {J.~I.}\ \bibnamefont {Gillen}},
  \bibinfo {author} {\bibfnamefont {A.}~\bibnamefont {Peng}}, \bibinfo {author}
  {\bibfnamefont {S.}~\bibnamefont {F\"{o}lling}}, \ and\ \bibinfo {author}
  {\bibfnamefont {M.}~\bibnamefont {Greiner}},\ }\href {\doibase
  10.1038/nature08482} {\bibfield  {journal} {\bibinfo  {journal} {Nature}\
  }\textbf {\bibinfo {volume} {462}},\ \bibinfo {pages} {74} (\bibinfo {year}
  {2009})}\BibitemShut {NoStop}%
\bibitem [{\citenamefont {Weitenberg}\ \emph {et~al.}(2011)\citenamefont
  {Weitenberg}, \citenamefont {Endres}, \citenamefont {Sherson}, \citenamefont
  {Cheneau}, \citenamefont {Schau\ss}, \citenamefont {Fukuhara}, \citenamefont
  {Bloch},\ and\ \citenamefont {Kuhr}}]{Weitenberg2011}%
  \BibitemOpen
  \bibfield  {author} {\bibinfo {author} {\bibfnamefont {C.}~\bibnamefont
  {Weitenberg}}, \bibinfo {author} {\bibfnamefont {M.}~\bibnamefont {Endres}},
  \bibinfo {author} {\bibfnamefont {J.~F.}\ \bibnamefont {Sherson}}, \bibinfo
  {author} {\bibfnamefont {M.}~\bibnamefont {Cheneau}}, \bibinfo {author}
  {\bibfnamefont {P.}~\bibnamefont {Schau\ss}}, \bibinfo {author}
  {\bibfnamefont {T.}~\bibnamefont {Fukuhara}}, \bibinfo {author}
  {\bibfnamefont {I.}~\bibnamefont {Bloch}}, \ and\ \bibinfo {author}
  {\bibfnamefont {S.}~\bibnamefont {Kuhr}},\ }\href {\doibase
  10.1038/nature09827} {\bibfield  {journal} {\bibinfo  {journal} {Nature}\
  }\textbf {\bibinfo {volume} {471}},\ \bibinfo {pages} {319} (\bibinfo {year}
  {2011})}\BibitemShut {NoStop}%
\bibitem [{\citenamefont {Endres}\ \emph {et~al.}(2011)\citenamefont {Endres},
  \citenamefont {Cheneau}, \citenamefont {Fukuhara}, \citenamefont
  {Weitenberg}, \citenamefont {Schauß}, \citenamefont {Gross}, \citenamefont
  {Mazza}, \citenamefont {Bañuls}, \citenamefont {Pollet}, \citenamefont
  {Bloch},\ and\ \citenamefont {Kuhr}}]{Endres2011}%
  \BibitemOpen
  \bibfield  {author} {\bibinfo {author} {\bibfnamefont {M.}~\bibnamefont
  {Endres}}, \bibinfo {author} {\bibfnamefont {M.}~\bibnamefont {Cheneau}},
  \bibinfo {author} {\bibfnamefont {T.}~\bibnamefont {Fukuhara}}, \bibinfo
  {author} {\bibfnamefont {C.}~\bibnamefont {Weitenberg}}, \bibinfo {author}
  {\bibfnamefont {P.}~\bibnamefont {Schauß}}, \bibinfo {author} {\bibfnamefont
  {C.}~\bibnamefont {Gross}}, \bibinfo {author} {\bibfnamefont
  {L.}~\bibnamefont {Mazza}}, \bibinfo {author} {\bibfnamefont {M.~C.}\
  \bibnamefont {Bañuls}}, \bibinfo {author} {\bibfnamefont {L.}~\bibnamefont
  {Pollet}}, \bibinfo {author} {\bibfnamefont {I.}~\bibnamefont {Bloch}}, \
  and\ \bibinfo {author} {\bibfnamefont {S.}~\bibnamefont {Kuhr}},\ }\href
  {\doibase 10.1126/science.1209284} {\bibfield  {journal} {\bibinfo  {journal}
  {Science}\ }\textbf {\bibinfo {volume} {334}},\ \bibinfo {pages} {200}
  (\bibinfo {year} {2011})}\BibitemShut {NoStop}%
\bibitem [{\citenamefont {{Neuenhahn}}\ and\ \citenamefont
  {{Marquardt}}(2010)}]{Neuenhahn2010}%
  \BibitemOpen
  \bibfield  {author} {\bibinfo {author} {\bibfnamefont {C.}~\bibnamefont
  {{Neuenhahn}}}\ and\ \bibinfo {author} {\bibfnamefont {F.}~\bibnamefont
  {{Marquardt}}},\ }\href@noop {} {\bibfield  {journal} {\bibinfo  {journal}
  {ArXiv e-prints}\ } (\bibinfo {year} {2010})},\ \Eprint
  {http://arxiv.org/abs/1007.5306} {arXiv:1007.5306 [cond-mat.stat-mech]}
  \BibitemShut {NoStop}%
\bibitem [{\citenamefont {Billingsley}(1995)}]{Billingsley}%
  \BibitemOpen
  \bibfield  {author} {\bibinfo {author} {\bibfnamefont {P.}~\bibnamefont
  {Billingsley}},\ }\href@noop {} {\emph {\bibinfo {title} {Probability and
  Measure}}}\ (\bibinfo  {publisher} {Wiley},\ \bibinfo {year}
  {1995})\BibitemShut {NoStop}%
\end{thebibliography}%

\end{document}